\documentclass[11pt]{article}
\usepackage[utf8]{inputenc}
\usepackage{times,fullpage,setspace}
\usepackage[english]{babel}
\usepackage[dvipsnames,table]{xcolor}
\usepackage{graphicx,mathtools,physics,amssymb}
\usepackage[margin=1in]{geometry}
\usepackage{multirow,url}
\usepackage{tikz,cite}
\usepackage{neuralnetwork}
   \usepackage{subfig}
\usetikzlibrary{arrows,positioning,chains}
\title{%
Applications of Artificial Intelligence, Machine Learning and related techniques for Computer Networking Systems \\ \ \\
}
\author{\textsf{Krishna M. Sivalingam, Professor} \\
\textit{Department of Computer Science and Engineering} \\ \textit{Indian Institute of Technology Madras, Chennai, INDIA} \\
Email: \texttt{\{skrishnam@cse.iitm.ac.in, krishna.sivalingam@gmail.com\}} \\
URL: \url{www.cse.iitm.ac.in/~skrishnam}
}
\date{\today}

\begin{document}

\maketitle

\begin{abstract}

This article presents a primer/overview of applications of Artificial
Intelligence and Machine Learning (AI/ML) techniques to address
problems in the domain of computer networking. In particular, the
techniques have been used to support efficient and accurate traffic
prediction, traffic classification, anomaly detection, network
management, network security, network resource allocation and
optimization, network scheduling algorithms, fault diagnosis and many
more such applications. The article first summarizes some of the key
networking concepts and a few representative machine learning
techniques and algorithms. The article then presents details regarding
the availability of data sets for networking applications and machine
learning software and toolkits for processing these data sets.
Highlights of some of the standards activities, pursued by ITU-T and
ETSI, which are related to AI/ML for networking, are also
presented. Finally, the article discusses a small set of
representative networking problems where AI/ML techniques have been
successfully applied.

\end{abstract}  

\section{Introduction}

Artificial Intelligence including Machine Learning (AI/ML) and related techniques have seen a substantial gain in several real-world applications in the past decade. In particular, AI/ML has been especially successful in  domains such as Image Processing, Computer Vision, Speech Recognition, Natural Language Processing, Robotics, Medicine, E-Commerce, Autonomous Driving, Security, Finance, and Transportation \cite{AIRussell2019,McKinsey2017,HBR2018,ExplAI2019}. 

We are at a crucial juncture where several machine learning (ML) and related techniques have matured over the past two decades and have been shown 
to be able to process large amounts of data. As mentioned above, ML has emerged as a crucial technology supporting the future of business, e-governance, entertainment, medicine, imaging and several other societal applications. In this paper, we present an overview of how AI/ML techniques have been and can be applied to the domain of Computer Networking.

The amount of data traffic carried by computer networks has grown exponentially in the past three decades. As per Cisco's VNI, the global Internet carries around.
Of this, around 70\% of the traffic is video-based, given the proliferation of e-learning classes, e-commerce, online meetings, online video exchange platforms, social media and messaging applications. The volume of data traffic grew significantly during the Covid-19 worldwide lockdown that started in early 2020 (and continues to last as of this publication). The network infrastructure is also getting complex, dynamic and adaptive. The backbone link capacities have correspondingly increased to several 100s of Gbps to a few Tbps, to keep pace the growing data traffic explosion. As a result, there is significant amounts of networking-based metadata that can be collected and analyzed for improving several aspects of a network such as network operation and management, network resource allocation, end-user quality-of-experience (QoE), quality-of-service (QoS) to traffic flows, and so on. 
The time has now come to effectively apply ML techniques in much larger scale. 

There have been several attempts to apply ML techniques to computer networking, starting in the early 2000s. These have shown significant promise. 
There have been a few survey articles on the application of AI and ML techniques for Computer Networks and specific domains within networks \cite{dqn,JISA:Surv18,Wang2018,ShenVTS20,KBN:CCR17,AI:OSN2018,Kafle2018,Bin2019,DataDriven5G2020,NetAI2020,IEEEAIWS2020}. 
What is now required is a giant leap forward to incorporate network data analytics and ML techniques in virtually all aspects of computer networking. The computer networking industry has also embraced this technology in a wholehearted manner, as seen from several product portfolios from leading hardware and software vendors, that have incorporated ML in their products. Standardization efforts have also gained major momentum as seen in the evolution of several ML-infused network architectures and frameworks, by international standards organizations such as 3GPP, ETSI and ITU-T \cite{eni,ENI:CommMag2018,ENI:CommMag2020,nwdaf,itut:y3172}. 

In this paper, we focus on how ML can be applied to the core network protocol layers and not on applications. There is significant amount of work done at the network-based application level, but is not considered here. Some of the main usage scenarios of ML in networks are: (i) Traffic Classification; (ii) Traffic Forecasting; (iii) Network Resource Management, (iv) Network Operations Control and Management; (v) Fault-tolerant Management including diagnostics and root-cause analysis, (vi) Protocol Design and Optimization, (vii) Data-driven network architecture design and (viii) Network Security, to name a few. This paper provides an overview of networking and key ML technologies, followed by a few case studies of recent uses of ML for computer networks.

The rest of the paper is organized as follows. 
Section~\ref{sec:nets} presents a primer to computer networking in order to explain the context in which ML techniques can be applied. Section~\ref{sec:ml} provides an overview of the typical ML tasks and their relation to computer networking activities. Section~\ref{sec:mltech} presents the details of some of the popular machine learning techniques used in recent times. Section~\ref{sec:mlsw} lists commonly used software packages and development toolkits with ready-made implementations of various ML algorithms. Section~\ref{sec:stds} briefly describes some of the standards activities related to the widespread adoption of ML for networks. Section~\ref{sec:uses} describes a few sample use cases where ML techniques have been successfully demonstrated for computer networking. Section~\ref{sec:concl} summarizes and concludes the paper.

\section{Computer Networking Primer}
\label{sec:nets}

This section presents a high-level overview of the important functionality and protocols used in computer networks. Specifically, the hardware and software components of a  networking system are discussed.

\begin{figure}[hbtp]
    \centering
    \includegraphics[width=0.9\textwidth]{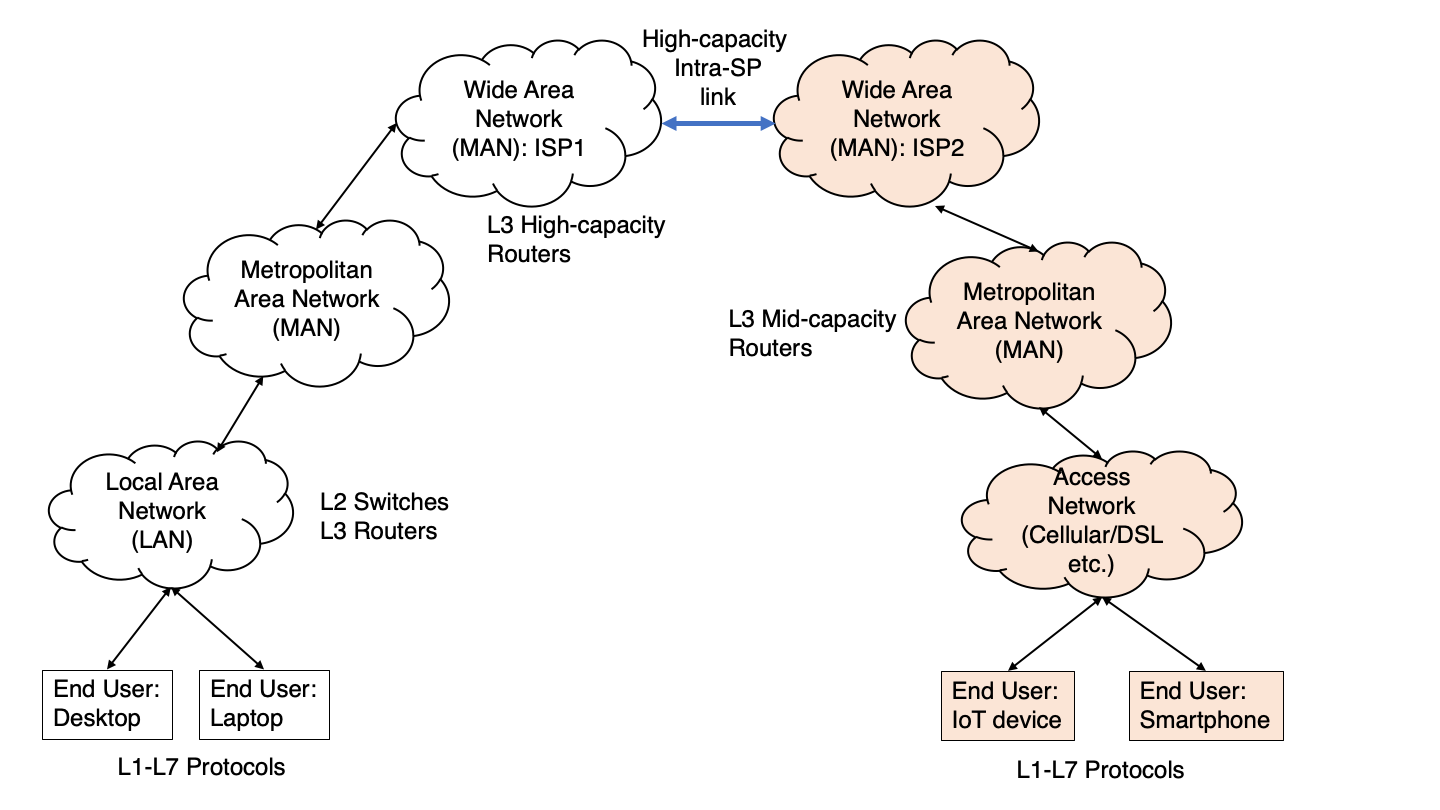}
    \caption{Network Architecture and Key Hardware Components.}
    \label{fig:cn}
\end{figure}

\subsection{Network Hardware}

Fig.~\ref{fig:cn} depicts a typical computer network architecture and
the various entities. The end-user equipment (UE) is typically a
desktop or laptop computer, tablet, mobile phone, Internet of Things
(IoT) device, etc. and in general refers to any device with a network
interface. The objective of the networking system is to interconnect
end users with each other. 

There are four major generic categories of networks: Access Networks, Local Area Networks (LANs), Metropolitan Area Networks (MANs) and Wide Area Networks (WANs). 
Access networks provide the proverbial last-mile connectivity to the users. Access Networks include Cellular Networks and Broadband Networks such as ISDN networks, Cable TV networks,  Digital Subscriber Line (DSL) networks and fiber optic networks. End-users connect to the Access Networks using devices such as modems, smartphones etc. and are connected to an Internet Service Provider (ISP). The access network bandwidths ranges are of the order of a few hundred Kbps (ISDN), tens of Mbps (DSL, Cable modems, 3G/4G Cellular networks etc.) and hundreds of Mbps to Gbps (5G networks, fiber optic networks). 

Local Area Networks are typically implemented in campus and enterprise environments. In these networks, end-users connect to the LAN using WiFi (IEEE 802.11), Ethernet (IEEE 802.3) or other similar LAN technologies using WiFi radios or Ethernet network interface cards. The networking equipment used in LANs are hubs, switches (e.g. Ethernet) and low- to mid-range capacity routers. A large enterprise network will have several interconnected LANs which are connected to the external network via specialized gateway router nodes, typically to an Internet Service Provider's network, called an Autonomous System (AS). Within the enterprise networks, there can be several other specialized network equipment for providing application servers, network security, improving performance and other purposes. These include firewalls, intrusion detection and prevention systems, email servers, file servers, web servers, database servers, and middleboxes that support Network Address Translation, application proxy, load balancing, security and protocol accelerators, and content caching.

The ISP networks primarily consist of Metropolitan Area Networks that cover a larger area, compared to a LAN, such as a city or large region. There will be  several such interconnected MANs covering a state or a province, using technologies such as SONET/SDH, OTN, or Carrier Ethernet. The MANs then feed into a large national backbone network that is realized using high-capacity backbone routers interconnected by ultra high-capacity (several hundreds of Gbps to few Tbps) links. 

The Internet, as we know it, is realized using an interconnected set of Autonomous Systems, with each AS potentially operated by a different ISP. Each AS with have a set of peer gateway nodes that connects to a subset of ASs using high-capacity links and specialized inter-AS routing protocols.

To summarize, the UE connects to the network using a wired network such as Ethernet, DSL or Cable Modem, etc or a wireless network such as cellular networks (2G -- 5G) or a wireless local area network (WLAN). This connection will typically be to an enterprise or campus network gateway router node or an Internet Service Provider (ISP). These in turn connect to the rest of the Internet via suitable Metropolitan and Wide Area Networks. 
The hardware entities that comprise the network include user equipment, LAN switches, routers, gateway nodes, firewalls, routers, MAN switches and routers, and WAN routers.

\bigskip
\begin{figure}[hbtp]
\rowcolors{2}{gray!25}{white}
\centering
\begin{tabular}{|l|c|c|c|c|}
\hline\hline
\rowcolor{gray!70}
\textbf{LAYER} & \textbf{NAME} & \multicolumn{3}{|c|}{\textbf{PROTOCOLS}} \\ \hline
\textit{Layer 7 } & \textbf{Application}  & {HTML} & Email & FTP  \\ \hline
\textit{Level 4.1} & \textbf{Security}  & \multicolumn{3}{|c|}{TLS}  \\ \hline
\textit{Level 4}  & \textbf{Transport}  & TCP & UDP & ICMP  \\  \hline
\textit{Level 3 } & \textbf{Network}  & \multicolumn{2}{|c|}{ IPv4} & IPv6\\ \hline
    &    &  \multicolumn{3}{|c|}{LLC} \\  \cline{3-5}
\multirow{-2}{*}{\textit{Level 2}} & \multirow{-2}{*}{\textbf{Link Control}} & \multicolumn{3}{|c|}{MAC}  \\  \hline
\textit{Level 1} & \textbf{Physical} & Copper & Fiber & Wireless \\  \hline
\hline
\end{tabular}
    \caption{TCP/IP Network Protocol Stack. Levels 5 and 6 that correspond to the seven-layer OSI stack are not presented here.}
    \label{fig:tcpip}
\end{figure}

\subsection{Network Software}

The required network software functionality is implemented in the end devices, switches and routers using a standard network software protocol stack that is based on the TCP/IP model or the OSI model. For this paper, we will refer to the commonly used TCP/IP model. Fig.~\ref{fig:tcpip} presents the layered protocol architecture. The topmost layer is the Application layer where the end-user application functionality is implemented. Popular application protocols support services such as Email, FTP, SSH, HTTPS, Social Media and so on. These applications are implemented in the user space of the Operating System running on the end device. The application programs contain the necessary application logic for realizing the end-to-end communication objectives, between two end-hosts.

In current generation secure networks, there is also a Transport Layer Security (TLS) protocol that resides between the application layer and the transport layer. The main functionalities of the TLS layer include end-to-end data encryption, mutual authentication between the end-hosts, message integrity and other related security aspects.

The next layer below is the Transport layer (also called Layer-4 or L4), which is  implemented in the kernel space of the Operating System. Here, Transmission Control Protocol (TCP), User Datagram Protocol (UDP) and Real-time Protocol (RTP) are commonly used protocols. 
TCP's main functionalities include end-to-end session establishment, reliable transmission of data segments, flow control and congestion control. TCP ensures that each data segment given by the application layer is successfully delivered to the receiving node's TCP entity, using sequence numbers, acknowledgments and retransmission of lost packets. Flow control is defined as a mechanism by which a fast sender cannot overwhelm a slow receiver, by slowing down the sender when necessary. Congestion control mechanisms are used to ensure that a TCP sender does not more traffic than the network can handle: when the network is congestion, TCP reduces the amount of traffic sent and vice versa. 
UDP, on the other hand, supports a simple datagram (individual message)  delivery service that does not provide reliable transmission. An application using UDP has to handle any  messages lost by the network. RTP is mainly used for supporting streaming applications such as video, teleconferencing and so on, and is not described here further, due to lack of space.

Another important functionality of transport layer protocols is multiplexing and demultiplexing of different application layer sessions over the underlying network layer interface. This is handled using the 16-bit port number field in the transport layer header. At the sender, the transport layer entity receives packets from different applications on their specified ports and forwards to the network layer. Similarly, at the receiver, the transport layer entity receives packets from
the the network layer and forwards the packets to the corresponding  application based on the specific ports the each application is 
registered for.

The main functionalities provided by the network layer (also called Layer-3 or L3),are addressing, routing and forwarding. Addressing refers to the naming of end-users in a network; the commonly used routing identifiers are hierarchical in nature such as the 32-bit IPv4 addresses (e.g. 128.205.31.1) and the 128-bit IPv6 identifiers. The global address space is administered by IANA (Internet Assigned Numbers Authority) that assigns globally unique network addresses to the different networks in the world. The unique end-user identifiers are assigned to end-nodes by the system administrators, either manually or using protocols such as Dynamic Host Configuration Protocol (DHCP). 

Routers are the main entities in the Internet that handle packet routing and forwarding.
The objective of routing is to forward a packet from the source node to the destination node via a series of intermediate routers that can span multiple ASs. Routing in the Internet is handled by two major type of protocols: intra-AS and inter-AS routing protocols. The former set of protocols are used to route packets between nodes within an AS, while the latter is used to route packets between nodes located in different ASs.  The commonly used intra-AS protocols include Open Shortest Path First (OSPF), Routing Information Protocol (RIP) and Information Systems-Information Systems (IS-IS) protocol, with Border Gateway Protocol (BGP) being a widely used inter-AS protocol. 

A router has multiple network interfaces. Forwarding within a router is defined as the action of identifying the appropriate network interface on which an incoming packet has to be transmitted on. This is done using forwarding tables in a router, that are populated using the various routing protocols mentioned above. 

The network layer also performs multiplexing and demultiplexing functions. Multiple transport layer protocols (such as TCP and UDP) and control protocols (such as Internet Control Message Protocol, ICMP) reside above the network layer and utilize its services. The network layer forwards packets received from the layer below (Layer 2) to the appropriate Layer-4 protocol. 

Note that the transport layer protocol software (TLS and TCP/UDP) is implemented only in the end-hosts in classic systems. Of late, some of the middleboxes also implement some of the transport layer functionality for efficiency and other reasons. In contrast, network layer software is implemented in both the end-hosts and in the network devices such as switches that support L3-functionality and in routers. In the end-host, the main functionality of routing software is that of multiplexing and de-multiplexing as mentioned above. In the routers and L3-enabled switches, the various routing protocols and related functionality are implemented.

The second layer in the protocol stack is the link layer (Layer-2 or L2). The link layer has two sub-layers, namely the logical link control (LLC) layer and the medium access control (MAC) layer. We will mainly discuss the MAC layer since it is more complex and has more potential of the application of machine learning techniques. The MAC layer is applicable when there is a medium shared among multiple contending users, as in shared Ethernet, IEEE 802.11 (WiFi) and in cellular network access. The MAC protocol determines which node(s) can transmit at a given point of time, and also handles collisions in case of multiple simultaneous transmissions. 

The link layer resides above and utilizes the services provided by the lowermost layer in the stack, which is called the physical layer (Layer-1 or L1). This layer handles the actual transmission of the data bits over the underlying medium such as wireless, copper, fiber, satellite channel, and so on.

Machine Learning can be potentially used in several of these hardware and software components, in order to provide enhanced network performance, improved user experience, efficient resource utilization, better protection from security attacks, reduced costs and in general, far more advanced next generation networking. 

\begin{figure}[hbtp]
\centering
\includegraphics[width=0.6\textwidth]{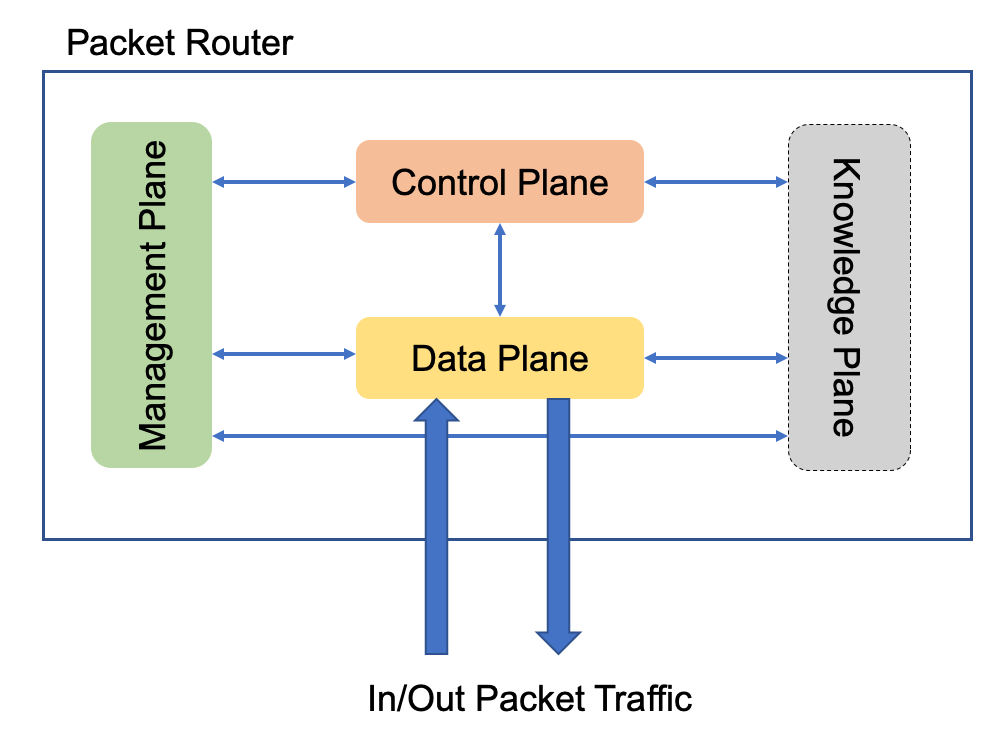}
\caption{Network planes, in a router.}
\label{fig:planes}
\end{figure}

\subsection{Network Planes}

The network functionality is also organized into three main planes: (i) Control Plane, (ii) Data Plane, and (iii) Management Plane. 
A router consists of multiple line cards (i.e. network interfaces) on which incoming packets (control, data and management) are received, processed and then forwarded onto the required outgoing interface. Fig.~\ref{fig:planes} shows the architecture of a typical router with the different planes.

The data plane primarily deals with receiving data packets on an input interface, processing the packets and forwarding data packets on the corresponding output interface. 
Data packets contain the information generated by the application layer (e.g. voice, video, ftp, http, ssh, etc.), along with the appended headers at the intermediate layers.  Data packets constitute a significant fraction (say 90\%) of the packets handled by a router. Packets will be queued in packet buffers at the output interface before being transmitted. The data plane deals with packet scheduling, active queue management,  quality-of-service implementation, congestion control and other related activities. The data plane implementation has to be exceptionally efficient in order to keep up with the very high incoming packet data rates (of the order of hundreds of Gbps). The data plane is realized using software and specialized hardware such as ASICs for efficient processing, and is run on the input and output line cards. Thus, the available data packet processing time is of the order of a few nanoseconds.

The control plane handles all packets related to control of network operations. A typical example is the routing protocol, which is used to determine paths or routes for a given destination node. For instance, the Open Shortest Path First (OSPF) routing protocol processes all the control packets that contain information about the network nodes and their current states, the interconnecting links and their current states and other details. Using OSPF packets received from all other routers in the network, a router constructs the routing table which will have the shortest paths from this router to all other destinations in this network or in an external network. 
Border Gateway Protocol (BGP), Routing Information Protocol (RIP) and Intermediate System to Intermediate System (IS-IS) protocol and Multi-Protocol Label Switching (MPLS) are examples of other routing protocols. Another important control plane protocol is the Internet Control Message Protocol (ICMP) that is mainly used for error reporting and network diagnostics.  The control plane is implemented in software and run on a special control plane card in the router; this card will contain a CPU, Main Memory and other necessary computing elements. The control plane packets have to be handled fast; however, there is more time available for processing such packets, of the order of microseconds.

The management plane deals with network element management, policy based configuration, monitoring, Service Level Agreements (SLA) compliance, fault diagnosis and handling, and other functionalities. This plane is implemented using a Network Management Systems (NMS) framework. Each network element (switch, router, firewall) will have a NMS agent (or sensor) that periodically records meta-data and other information about all packets flowing through the device. This is shared on a regular basis through a hierarchy of such agents with a centralized NMS server. Thus, a substantial amount of operational data is processed by the server, which constantly monitors the system and key performance metrics for identifying any issues such as poor performance, high utilization, service disruptions, high packet buffering and queuing delays and so on.  The NMS framework is a natural place for machine learning algorithms to be implemented. The framework will also issue necessary alerts for any correction action to be taken autonomously by the network elements or in conjunction with the practical expertise of system administrators. Though the volume of data handled by the NMS system is very high, the processing time requirements are relatively relaxed compared to the data  and control planes, and are of the order of milliseconds to a few seconds. However, during network failure events, the timing requirements to detect and correct failures will be quite stringent especially in multi-Terabit capacity networks. Simple Network Management Protocol (SNMP) is the most widely used management plane protocol.

In addition, with the growth and potential of machine learning and analytics applications for computer networks, a fourth plane called the \textit{Knowledge Plane} has been proposed \cite{KnowPlaNet2003}. This plane was conceived to be based on cognitive techniques with the associated knowledge representation, behavioral models, learning and reasoning to be aware of the network state and perform necessary actions. The potential applications of this plane, as described in \cite{KnowPlaNet2003}, include fault management, dynamic and automatic (re)configuration of network elements, and intrusion detection. The current generation systems have not fully achieved such a conceptual vision, but are being designed to achieve similar goals for the next generation of networks.
Recently, the work on \textit{Knowledge-Based Networking} \cite{KBN:CCR17} presents a network architecture that combines software defined networking principles, machine learning techniques and analytics platforms to realize closed-loop and open-loop control of computer networks. The objectives are to provide "provide automation, recommendation, optimization, validation and estimation" \cite{KBN:CCR17}.

\section{Machine Learning Primer}
\label{sec:ml}

This section presents some of the common machine learning (ML) tasks. 
ML algorithms are classified into \textit{supervised}, \textit{unsupervised}   and \textit{semi-supervised} learning algorithms.

\subsection{Supervised Learning}

The inputs to a supervised machine learning algorithm typically include a set of $m$ samples, where each sample has a set of features, denoted as $\vb{x} = \{x_1, x_2, ..., x_n\}$, where $x_i$ denotes the $i^{th}$ feature. The range of values for each $x_i$ can be different. There is also an output variable that depends on the values of $x_i$, denoted as $y = f(x_1, x_2, ..., x_n)$, where the function $f$ is unknown. Thus, for each sample $i$, the values of the input variables ($\vb{x}^{(i)}$) and the output variable ($y^{(i)}$) are known. This value of $y^{(i)}$ is sometimes referred to  as the ``label'' or ``ground truth'' of sample $i$. Given  $m$ samples of $\{(\vb{x}^{(i)}, y^{(i)})\}$ values, the  objective of the machine learning algorithm is to approximate the function $f(\vb{x})$ using a hypothesis, denoted as  $h(\vb{x})$. Then, given a new input $\vb{x}$, the corresponding predicted value of $y$ can be obtained using $h(\vb{x})$.

If $y$ is a continuous real variable, then this formulation is referred to as a \textit{regression} or \textit{prediction} problem. 
For example, consider a set of routers in a ISP's network. Each router reports the number of packets, the number of bytes
and the queue length of a specific outgoing interface along with the latency per packet, once an hour, to a central system. These represent the $\vb{x}^{(i)}$ values of a given router. Also, each router measures and reports the average packet latency measured over the packets handled during the specified hour. The machine learning algorithm run at the central system can then learn from  samples (taken over a 3-month period) and predict the average packet latency given a new sample value of $\vb{x}$. 

If $y$ takes on values from a discrete set, then this is referred to as the \textit{classification} problem. For example, consider a backbone router that measures the number of packets, the inter-packet arrival time and the number of bytes handled for each flow passing through the router. Assume that the flows are labeled as belonging to one of four classes, namely \textit{email}, \textit{ftp}, \textit{http} and \textit{ssh}. Given a set of labeled flows, the machine learning algorithm can be used to classify a new flow for which the set of features mentioned above are collected. This classification can be then used by the router to decide upon suitable forwarding, buffering and queuing strategies to handle the different flow classes. 

One of the common challenges in classification problems is that the labelled values can be unbalanced. That is, it is possible to have substantially more samples of one label compared to the others. For example, there may be several examples of HTTP packets in a dataset compared to say, SSH packets, in a packet classifier. In such cases, the model parameters for classifying SSH might not be accurate. Thus, there is a need to design classifiers that are able to handle such data imbalance without bias, and still classify accurately \cite{ImbClass2009}. 

\subsection{Unsupervised Learning}

The inputs to an unsupervised machine learning algorithm  include a set of $m$ samples, where each sample has a set of features, denoted as $\vb{x} = \{x_1, x_2, ..., x_n\}$, where $x_i$ denotes the $i^{th}$ feature, as earlier.  However, unlike supervised learning, there is no specific output value or label assigned to a sample. The learning algorithm has to learn about the structure of the data by understanding the relationships among the different features. 

Some of the common tasks where unsupervised learning is used are clustering, dimensionality reduction and density reduction. With clustering, the learning algorithm tries to cluster or segment the different data samples into various groups, based on some similarity measure among the data samples. The number of clusters is not always known in advance and is specified as an additional input to the ML algorithm. 

For example, consider an intrusion detection system (IDS) deployed in a network; the IDS monitors the traffic entering the network from the outside and performs the task of anomaly detection. Here, the ML model has to learn what is ``normal'' network traffic and raise an alarm whenever it observes that there is an anomaly, i.e. significant deviation from normal behavior. For instance, this could be a case where there is a significant increase in packets traversing the network, that could be due to a virus attack triggered in the network. In this case, it is not possible to label the network traffic characteristics ahead of time and the ML system has to learn how to predict anomalous behavior.

In dimensionality reduction (DR), the algorithm tries to determine the most important features of the inputs and reduces the number of features that is fed to another subsequent learning algorithm. For example, consider a scenario where a network management system collects 100 different network and router management variables, of which only 20 have a direct impact on system performance.  DR helps determine the key features for further learning purposes. This can help in reducing the complexity of the original learning problem.

\subsection{Semi-supervised Learning}

Semi-supervised learning algorithms have a large data set, of which only a small number of samples are labeled. The ML algorithm learns using these labels and this model is applied to the unlabeled samples in order to label them. This fully-labeled data set is then used for training the ML model, that is then used to label new unlabeled samples provided to the model. For example, consider a spam detection service provided by an email provider. The provider can collect a large set of sample emails, of which a small subset can be manually labeled as spam and non-spam emails. The ML algorithm trains itself using the small labeled set and then used this model to label the remaining unlabeled email samples. The resultant model is then used against new emails for detecting spam.

\subsection{Reinforcement Learning}

Reinforcement Learning (RL) is an agent-based sequential decision-making technique in which the system agent interacts with the environment. The system behavior is captured as a\textit{ }series of discrete steps.  In each step, an action ($a$) is taken by the agent based on the current system \textit{state} ($s$), from a set of possible actions in that state. The strategy used by the agent in determining the action in a given state is called the \textit{policy}. This action results in the system state changing to a new state ($s'$). There is a \textit{reward} (or penalty) associated with action $a$ taken by the system in a given state. The system model is updated based on the action taken and the corresponding reward, at each step. The overall objective of the system is to maximize the accumulated reward over a period of time in order to achieve a specific of goals. This can be considered as a \textit{trial-and-error} based learning technique, where the agent combines \textit{exploration} (trying out new actions that have not been tried earlier) and \textit{exploitation} (selecting actions based on its earlier experiences).

RL has been successfully applied in several application domains such as robotics, autonomous driving, intelligent game playing (Chess, Go, etc.) and resource management in networks and in cloud systems. One of the major disadvantages of RL is scalability, making it unsuitable for highly complex systems, where one encounters the combinatorial explosion problem due to a large number of possible states and actions. Deep Neural Networks (DNN), discussed later, have been effectively combined with RL techniques to realize Deep Reinforcement Learning (DRL) techniques that can significantly mitigate the scalability concerns \cite{drl2018}.

\section{Common Machine Learning Techniques}
\label{sec:mltech}
This section explains some of the commonly used machine learning techniques, specifically for computer networking data. There are several other techniques that are available, but are not listed here for brevity.

Notation: Let $\vb{x}$ denote a column vector with the following elements, $x_1, x_2, \ldots{}, x_n$, and $\vb{x}^T$, the transpose of $\vb{x}$. The notations are similar to those used in \cite{CourseraNgML}.

\subsection{Linear Regression} Linear Regression is a classic technique that is used to predict the variable $y$ as a linear function of the various input $\vb{x}$. Given a set of $m$ samples denoted by $\vb{x}^{(i)}, i \in \{1, 2, ..., m\}$, the linear regression algorithm obtains the values of the parameters (or coefficients or weights), denoted by $\vb{\Theta} = \{\theta_0, \theta_1, ..., \theta_n\}$, where $\theta_0$ the baseline weight.  This is done by minimizing a suitable cost function such as  the \textit{mean squared error (MSE)} using an algorithm such as \textit{gradient descent}.  The predicted value is given by:

\begin{align}
\hat{y} = h(\vb{x}) = \Theta^T \vb{x} = \sum_{i=0}^{n} \theta_i x_i = \theta_0 x_0 + \theta_1 x_1 + ... + \theta_n x_n 
\end{align}

Here, $\hat{y}$ denotes the predicted output value, $h$ the hypothesis function determined based on the input samples, and $x_0$ equals $1$.
Consider the following $m=10$ sample data points, with two features/variables, $x_1$ and $x_2$, and the corresponding values of $y$.

\begin{center}

\begin{tabular}{|c||c|c|c|c|c|c|c|c|c|c|}
\hline
\textbf{Sample} & 1 & 2 & 3 & 4 & 5 & 6 & 7 & 8 & 9 & 10 \\ 
\hline \hline
 $x_1$  &  0 & 2 & 4 & 6 & 8 & 10 & 12 & 13 & 15 & 17 \\ \hline
  $x_2$    & 2 & 4 & 5 & 9 & 5 & 13 & 15 & 21 & 23 & 16  \\ \hline \hline
   $y$ & 4 & 8 & 11 & 9 & 13 & 24 & 30 & 19 & 21 & 24 \\ \hline
\end{tabular}
    
\end{center}

Using the MSE cost function, the corresponding coefficients are determined to be $\Theta = \{ 5.265, 1.335, -0.051 \}$. For new input data, $\vb{x} = \{1, 4\}$, $y = 6.395$; and $\vb{x} = \{14, 12\}$, $y = 23.339$.

The linear regression method can be extended to polynomial regression, where the variables $x_i$ can also be combined to create higher-order hypothesis. For example, consider a scenario with 3 input features ($x_1, x_2, x_3$); a polynomial hypothesis function of degree 2 can defined as:

\[\hat{y} = h(\vb{x}) = \theta_0 x_0 + \theta_1 x_1 + \theta_2 x_2 + \theta_3 x_1 x_2 + \theta_4 x_1 x_3 + \theta_5 x_2^2\]

For a different scenario with 2 input features ($x_1, x_2$), a polynomial hypothesis function of degree 3 can defined as:

\[\hat{y} = h(\vb{x}) = \theta_0 x_0 + \theta_1 x_1 + \theta_2 x_2 + \theta_3 x_1 x_2 + \theta_4 x_1^2 x_2 + \theta_5 x_1 x_2^2 +  \theta_6 x_1^3 + \theta_7 x_2^3\]

The polynomials can be designed for the desired degree, $d$. In general,  determining a suitable polynomial is part of the data analyst's expertise and can be obtained using some combination of prior domain knowledge and trial-and-error attempts. The choice of the hypothesis function depends on the understanding of the relationships among the different features and the output value.

\subsection{Logistic Regression}

The logistic regression technique is used for classification purposes. Consider a simple binary classifier that classifies emails as Spam or Not-Spam. The inputs to the classifier are email texts that have been manually labeled as Spam or otherwise; assume that $y = 0$ indicates Spam and $y = 1$ indicates Not-Spam. The hypothesis function is determined based on the sample inputs and is given by:

\begin{align}
    h_\Theta(\vb{x}) = g(\Theta^T \vb{x}) \\
    g(z) = \frac{1}{1 + e^{-z}} 
\end{align}

Here, $g(z)$ denotes the \textit{sigmoid} or \textit{logistic} function and is defined as follows:
$g(z) \geq 0.5, \mathrm{if } z \geq 0$ and $g(z) < 0.5, \mathrm{if } z < 0$. The vector $\Theta$ is learned by the algorithm based on the provided input samples. 
Thus, the hypothesis function returns the value of $y = 1$ if $\Theta^T \vb{x} \geq 0$ and $y = 0$, otherwise. 

Logistic Regression provides the decision boundary that can be used to determine the appropriate class. This boundary can be linear or non-linear. Consider a system with two features or variables, denoted by $x_1$ and $x_2$. A linear decision boundary is given by:

\[\theta_0 x_0 + \theta_1 x_1 +  \theta_2 x_2, \mathrm{where\ } x_0 = 1 \]

An example non-linear decision boundary is given by:

\[\theta_0 x_0 + \theta_1 x_1 +  \theta_2 x_2^2 + \theta_3 x_1^2 x_2^3, \mathrm{where\ } x_0 = 1 \]

The binary classifier can be extended to a multi-class classifier using different methods, one of which is the ``one-versus-all'' method. Here, for $\vb{C} = \{C_1, C_2, ..., C_k\} $ classes, the binary classifier is run $k$ times, since there are $k$ classes. For each run, the samples belonging to class $C_i$ are labeled $y = 0$ and all other samples are labeled as $y=1$ to obtain the hypothesis function for that class, denoted by $h_\theta^{(i)}(\vb{x})$. This represents the probability that $y=i$ for the given input. The training procedure is repeated for all values of $i \in \{1, 2, ..., k\}$ to obtain the corresponding hypothesis functions. 
Given a test input $\vb{x}$, the outputs of all the hypothesis functions are obtained. The hypothesis function that yields the highest value is used to classify the input as belonging to the corresponding class, i.e. Class~$\displaystyle\operatorname*{arg\,max}_i h_\theta^{(i)}(\vb{x})$.

An alternative method, called ``one-versus-one'' is based on constructing a set of pairwise classifiers, but is not considered further here.

\subsection{Support Vector Machines (SVM)}

The SVM technique is similar to logistic regression 
and is used for classification problems in machine learning.
SVM determines the decision boundary that separates the different data points in the different classes, such that the distance between the boundary and the closest element of each class is maximized. It is also referred to as a \textit{large margin} classifier due to this reason. We will consider a binary classifier in the following discussion,

SVM can deal with both linear (for example, a straight line in a 2-dimensional feature space) and non-linear decision boundaries. SVM uses the concept of a \textit{kernel} to specify the computation. A kernel is a specially designed function that maps some of the original features into new features, followed by learning the model using these new features. If no kernel is used, it is referred to as SVM with a linear kernel. An example of a non-linear kernel is the Gaussian kernel given by:

\begin{align}
 \displaystyle   f_i = e^{-\frac{ \norm{\vb{x} - L^{(i)}}^2}{2 \sigma^2}}
\end{align}

where a set of points $L^{(i)}, i = 1, 2, ..., k$, called landmark points, are chosen in the given feature space. This results in creation of $k$ new features, namely $f_{i}, i = 1, 2, ..., k$. The hypothesis function is defined by $\Theta_T f$, which is used to predict if $f = 0$ or $f=1$.

Several other kernels are possible, such as the polynomial kernel, defined by $(\vb{x}^T l + c)^d$, where $c$ denotes a constant and $d$ represents the chosen degree of the polynomial.

For multi-class classification, the binary SVM classifier can be extended using the ``one-vs-all'' or the ``one-vs-one'' approaches. 
SVM can also be used for the regression task, when is referred to as the Support Vector Regression (SVR) technique.

\subsection{K-Nearest Neighbors (KNN) Algorithm}

The K-Nearest Neighbors (KNN) algorithm is another popular supervised learning method used for both regression and for classification tasks. Here, $K$ is an input to the algorithm along with the labeled samples from the data set. In this method, given an unclassified input $\vb{x}$, its distance (as defined by an appropriate function) to each node in the sample data set is calculated. For instance,  Euclidean distance can be used to calculate the distance between the points. The samples are then sorted based on the calculated distances and the top $K$ nodes in this sorted list are selected. For regression, the mean of the corresponding ($y$) values of these $K$ samples is returned as the output. For classification, the mode of the output ($y$) labels of these $K$ samples  is returned as the output. It is a simple algorithm, but will be computationally expensive for very large data sets.

\subsection{K-Means Clustering}

This is a classic unsupervised learning algorithm that is used for clustering data into different groups or segments. The input to the algorithm is the training set consisting of $m$ samples, where each sample is an n-dimensional data point, and $K$, the number of desired clusters. 

The naive K-means algorithm chooses $K$ randomly selected centroids (could be a subset of the sample data too), and assigns each data point to the closest centroid in an iterative manner; at each step of the iteration, the centroid of each cluster is also re-calculated based on the data points assigned to the cluster. This process is repeated until the assignments of data points to clusters does not change. 

In practice, the algorithm is run multiple times (say, 100) with different random initial centroids.  Then, the clustering assignment that minimizes
a specific optimization objective is selected as the final result. Let the clusters be denoted by $\vb{S} = \{S_1, S_2, ..., S_K\}$, with corresponding centroids denoted as $\{\mu_1, \mu_2, ..., \mu_k$. The training samples are denoted by $\vb{x}^{(i)}$ as earlier. One optimization objective based on within-cluster sum of squares (WCSS) is given by:

\begin{align}
   \operatorname*{arg\,min}_{\vb{S}} \sum_{i=1}^{K} \sum_{\vb{x} \in \vb{S}_i } \norm{\vb{x}-\mu_i}^2
\end{align}

There are more efficient algorithms that can run much faster, but are not discussed here.

\subsection{Principal Component Analysis (PCA)}

Principal Component Analysis (PCA) is a widely used technique for dimensionality reduction, and is an unsupervised learning algorithm. It is commonly used for data compression and enhanced data visualization.

Consider a data set with $m$ samples, where each sample is a $n$-bit vector, denoted by $\vb{x} = \{x_1, x_2, ..., x_n\}$. PCA can reduce the feature space from $n$-dimensions to $k$-dimensions, where $k \ll n$. 
The algorithm computes the co-variance matrix of the $n$ features. The $\mathrm{CoV}(x_1, x_2)$ denotes the correlation between these two features; if this value is zero, then the 2 features are not correlated, but not necessarily independent of each other. If the covariance is positive, then the two features will increase or decrease in the same direction; if the covariance is negative, then the two features will vary in opposite directions. 

Next, the eigenvalues and eigenvectors of the co-variance matrix are calculated. Then, the $k$ features with the highest eigenvalues are  selected for further learning purposes. These features will contribute the maximum to the total variance and are hence selected. The choice of $k$ is up to the application developer; however, it can  also be obtained based on a target value of percentage of variance retained. 

\subsection{Decision Tree and Random Forest}

The Decision Tree technique can be used for supervised classification and regression. We will consider classification first. The inputs to the technique are the data samples along with the corresponding labels. The technique constructs a (typically) binary decision tree based on rules inferred from the the data. Each node in the tree contains a set of rules (specific value ranges for a subset of the data features) and in some cases, the corresponding value. Given the tree and an unclassified input, the tree is traversed from the root onwards by comparing the input's features against the rules stored in each node of the tree. At the end of the traversal, the class of the input is determined.

\begin{figure}[bhtp]
    \centering
    \newdimen\nodeDist
\nodeDist=35mm
\begin{tikzpicture}[
    node/.style={%
      draw,
      rectangle,
    },
  ]

    \node [node] (A) {\textit{Inter-arrival time $>$ 1~ms}?};
    \path (A) ++(-135:\nodeDist) node [node] (B) {\textbf{Mice Flow}};
    \path (A) ++(-45:\nodeDist) node [node] (C) {\textit{ByteCount $>$  10~MBytes?}};
    \path (C) ++(-135:\nodeDist) node [node] (D) {\textbf{Mice Flow}};
    \path (C) ++(-45:\nodeDist) node [node] (E) {\textbf{Elephant Flow}};

    \draw (A) -- (B) node [left,pos=0.25] {no}(A);
    \draw (A) -- (C) node [right,pos=0.25] {yes}(A);
    \draw (C) -- (D) node [left,pos=0.25] {no}(A);
    \draw (C) -- (E) node [right,pos=0.25] {yes}(A);
\end{tikzpicture}

    \caption{Flow Classification based on packet statistics of the flow.}
    \label{fig:dt}
\end{figure}
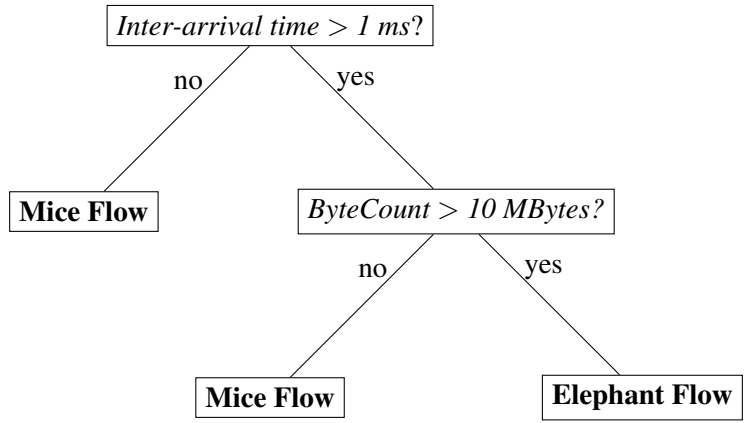

Fig.~\ref{fig:dt} presents an example decision tree for packet flow type classification based on two features, namely  inter-arrival time and bytes transmitted. The optimal tree construction problem, for a given set of inputs, is an NP-complete problem. Some of the commonly used algorithms for tree construction are Iterative Dichotomiser (ID3) and C4.5. 

The decision tree technique does not require data pre-processing such as feature scaling or normalization, which is often recommended for other ML techniques. Further, the tree can be easily visualized and thus the technique is considered as an "explainable" algorithm. Also, the classifier can be used for prediction purposes too and in particular, it can predict multiple output variables. For instance, the  Classification and Regression Trees (CART) algorithm can be used for regression tasks.

A disadvantage of the decision tree technique is that it can be sensitive to small variations in the input data, leading to the construction of a different tree, in terms of the orthogonal boundaries used in the decision making. One way to improve the performance of decision tree is by constructing an \textit{ensemble} of several decision trees to create a \textit{Random Forest}. Here, the different decision trees are constructed for the same input, by introducing some form of randomness in the tree construction process. Given an unclassified input, the decision tree outputs of all the trees in the forest are averaged to determine the final class of the input.

There are several other ensemble methods based on constructing multiple decision trees, such as Bagging, AdaBoost, Gradient Tree Boosting, XGBoost \cite{xgboost} and so on. These are not discussed here for the sake of brevity.

\subsection{Time Series Analysis}

The Time Series Analysis and Time Series Forecasting (TSF) techniques deal with forecasting future trends based on data collected at regular time intervals. For example, the daily closing price of a given stock can be collected over a time period (say, 5 years) and the future stock value can be predicted. Other examples include weather forecasting based on past weather data and predicting future sales of a product. In computer networks, TSF can be used to predict the network traffic at a router, to predict the queue lengths in a router's ports, predict link utilization in a network, and so on.

TSF belongs to the general class of Time Series Modeling. To perform TSM, the underlying distribution of the time series data should satisfy the \textit{stationarity} properties: the mean and variance should be a constant and should not vary with time; likewise, the covariance between the $i^{th}$ and $(i+m)^{th}$ sample should be a constant and should not vary with time. If a series is not stationary, it can be converted to meet the stationary properties.

Some of the common TSM models are based on autoregressive–moving-average (ARMA), AutoRegressive Integrated Moving Average (ARIMA),  ARCH (autoregressive conditional heteroskedasticity) and the GARCH (generalized autoregressive conditional heteroskedasticity).

\subsection{Neural Networks}

A Neural Network is a classical machine learning technique that is based on modeling the computation similar to how the human brain works. The brain processes various signals and achieves required tasks, using neurons, synapses,  neurotransmitters and electric signals. Neural networks attempt to vaguely mimic this behavior by having nodes that represent neurons, arranged is a series of layers and using weighted sums of the node values from one layer to be used for computations in subsequent layers. They have been used especially as non-linear function approximators and are particularly useful where linear regression fails.

As defined earlier, consider set of $m$ samples, where each sample has a set of features, denoted as $\vb{x} = \{x_1, x_2, ..., x_n\}$, where $x_i$ denotes the $i^{th}$ feature. There can be multiple output variables that depends on the values of $x_i$, denoted as $y_j = f_j(x_1, x_2, ..., x_n), j = 1, 2, 3, ..., F$, where the functions $f_j$ are unknown.

\begin{figure}
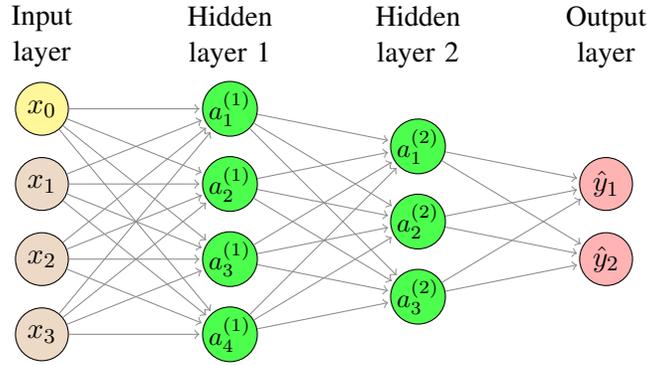

    \centering
    \begin{neuralnetwork}[height=4]
     \tikzstyle{input neuron}=[neuron, draw=black, minimum size=20pt, fill=brown!30];
  \tikzstyle{output neuron}=[neuron, draw=black, minimum size=20pt,fill=red!30];
  \tikzstyle{hidden neuron}=[neuron, draw=black, minimum size=20pt, fill=green!70];
  \tikzstyle{bias neuron}=[neuron, draw=black, minimum size=20pt, fill=yellow!50];
        \newcommand{\x}[2]{$x_#2$}
        \newcommand{\y}[2]{$\hat{y}_#2$}
        \newcommand{\hfirst}[2]{\small $a^{(1)}_#2$}
        \newcommand{\hsecond}[2]{\small $a^{(2)}_#2$}
        \inputlayer[count=3, bias=true, title=Input\\layer, text=\x]
        \hiddenlayer[count=4, bias=false, title=Hidden\\layer 1, text=\hfirst] \linklayers
        \hiddenlayer[count=3, bias=false, title=Hidden\\layer 2, text=\hsecond] \linklayers
        \outputlayer[count=2, title=Output\\layer, text=\y] \linklayers
    \end{neuralnetwork}
\caption{An example neural network with 2 hidden layers and two output functions.}
\label{fig:nn}
    
\end{figure}

The Multi-layer Perceptron (MLP) is one such supervised learning model. As shown in Fig.~\ref{fig:nn}, the model consists of an input layer (that represents the $n$ features of the input data) and an output layer (that represents the $F$ functions to be determined). In between these two layers, there exist several intermediate \textit{hidden} layers; interconnections are defined between successive layers, and each interconnection is assigned a specific weight. In the diagram, there are 2 output variables and 2 hidden layers. Given a specific sample, its feature values are assigned to the first layer, with $x_0 = 1$. For every other node, the value of the node is a linear sum of the corresponding previous layer's nodes that are connected to it and the weights on the corresponding links. 
The objective of the neural network training algorithm is to determine these weights based on the input data, such that the outputs for an unknown input can be determined using the approximate functions, $\hat{f}_j$. MLPs can be used for both regression and classification purposes. 

The specific number of hidden layers, the number of nodes in each hidden layer and the interconnections between the layers are determined by the domain expert. The example shown above is classified as the feed-forward neural network; other types such as convolutional neural networks (CNNs), graph neural networks (GNN), graph convolutional networks (GCN), and recurrent neural networks (RNNs) also exist, but are not described here.

Neural networks tend to be computationally more intensive and have been widely used for image processing, vision and NLP applications. After a lull in the 1990s, the availability of high-performance computing and vector computing systems (such as GPUs), neural networks and especially deep neural networks became very important machine learning tools for addressing several challenging learning tasks. There has been considerable interest in applying DL techniques to addressing computer networking problems.

Neural networks, where the number of layers is three or higher with potentially a very large number of neurons per layer, are called \textit{deep neural networks} (DNNs). Machine learning using DNNs is referred to as \textit{deep learning (DL)} \cite{Goodfellow2016}. 
DNNs are typically \textit{feedforward} networks, where computation flows from the input layer to the output layer. A class of DNNs is called recurrent neural networks (RNN), where feedback loops are included in the feedforward network, and useful for processing sequential data. One such network is the Long Short-Term Memory (LSTM) model, and are particularly suited for handling time series data. In the context of computer networking, LSTM models have been used for anomaly detection in network traffic, intrusion detection and to predict future network traffic patterns.

\subsection{Markov Decision Processes and Reinforcement Learning}

The Reinforcement Learning (RL) process is formally represented using a discrete-time stochastic process. Let $\mathcal{S}$ and  $\mathcal{A}$  denote the set of states in the learning environment and the set of all possible actions that the agent can take, respectively. For example, consider a $4 \times 4$ grid on which a robot moves around in a room. Each square cell $\mathcal{S} = \{(i,j) | i, j \in \{0, 1, 2, 3\} \}$ in the grid denotes a possible state for the robot to be in.
The possible actions that the robot can take denotes its movement to an immediate neighboring cell; assume that the set of  actions is represented by $\mathcal{A} = \{U, R, D, L\}$, denoting up, right, down and left movements.

Let the agent start from an initial state, $s_0 \in \mathcal{S}$, with the agent making a system observation denoted by $w_0 \in \mathcal{W}$. The state of the system at  time step $t$ is denoted by $s_t$. Let $a_t$ denote the action taken by the agent at time step $t$. As a result, the agent state changes to $s_{t+1}$, the agent earns a reward $r_t \in \mathcal{R}$, and the agent makes an observation denoted by $w_{t+1} \in \mathcal{W}$. 

A special case of the discrete-time stochastic process is the Markov Decision Process (MDP) defined by the 5-tuple $\mathcal{S}, \mathcal{A}, T, R, \gamma$. Here, $T: \mathcal{S} \times \mathcal{A} \times \mathcal{S} \rightarrow [0,1]$ denotes the transition probabilities between the different system states. $R: \mathcal{S} \times \mathcal{A} \times \mathcal{S} \rightarrow \mathcal{R}$ denotes the reward function, and $\gamma \in [0, 1)$ denotes a discount factor for past rewards. In an MDP, the system is fully observable, i.e. $w_t = s_t$. There are special cases where the system is partially observable, but those are not considered here. At each time instant $t$, the probability of the system state changing to $s_t+1$ is given by $T(s_t, a_t, s_t+1)$ and the reward earned is given by $R(s_t, a_t, s_t+1)$. 

The objective of the learning algorithm is to learn the policy ($\pi \in \Pi$) that specifies the process by which an agent selects the action to be taken at a given instant of time. Here $\Pi$ denotes the set of all feasible policies. A policy is classified as \textit{stationary} if it does not depend on the current time step, i.e. it does not change with time. A policy is classified as \textit{deterministic}, where the action taken for a given system state is fixed by the policy. This is denoted by  $\pi(s): \mathcal{S} \rightarrow \mathcal{A}$. Alternatively, the policy can be stochastic, where an  action $a$ taken in a given state $s$ with a specified probability. This is denoted by  $\pi(s, a): \mathcal{S} \times \mathcal{A} \rightarrow [0, 1]$.

For the deterministic policy, the objective is thus to find an optimal policy ($\pi^*(s)$) that optimizes an expected discounted reward function, called the V-value function, defined as:

\begin{align}
V^\pi(s) &= \mathcal{S} \rightarrow \mathcal{R} \\
V^\pi(s) &= \left[ \mathbb{E} \sum_{k=0}^{\infty} \gamma^k r_{t+k} | s_t = s, \pi \right] \\
    V^*(s) &= \max_{\pi \in \Pi} V^\pi(s) \\
    \pi^*(s) &= \operatorname*{arg\,max}_{\pi \in \Pi} V^\pi(s)
\end{align}

Thus, $V^*(s)$ is the  expected discounted reward when the agent follows the optimal policy $\pi^*(s)$, from a given state $s$.

Similarly, for the stochastic policy, the objective is thus to find an optimal policy ($\pi^*(s)$) that optimizes an expected discounted reward function, called the Q-value function, defined as:

\begin{align}
Q^\pi(s,a) &= \mathcal{S} \times \mathcal{A} \rightarrow \mathcal{R} \\
Q^\pi(s, a) &= \left[ \mathbb{E} \sum_{k=0}^{\infty} \gamma^k r_{t+k} | s_t = s, a_t = a, \pi \right] \\
    Q^*(s,a) &= \max_{\pi \in \Pi} Q^\pi(s,a) \\
    \pi^*(s) &= \operatorname*{arg\,max}_{a \in \mathcal{A}} Q^*(s,a)
\end{align}

Thus, $Q^*(s,a)$ is the  expected discounted reward when the agent follows the optimal policy $\pi^*(s)$, from a given state $s$ and performing an action $a$.

Learning the optimal policy can be done in an offline setting (or batch setting), where a set of limited data is provided for learning. The alternative is the online setting, where the agent interacts with the environment and uses the experiential information for learning the policy. In the context of networking, online systems seem to be preferable.

The computational methods that obtain the optimal policy are discussed next. One simple method is to use a Monte Carlo simulation of the system and run it for a very long run with multiple random seeds and obtain the optimal policy. However, it is computationally inefficient, especially for complex systems. A simple and popular algorithm, that is also efficient in terms of computation, is the Q-learning algorithm \cite{Qlearn}. Other algorithms are Q-fitting \cite{Qfit96} and neural fitted Q-learning (NFQ) \cite{nfq}.

One of the challenges with RL algorithms is the explosion of parameter space as the system complexity increases. Deep learning approaches combined with RL led to the design of Deep Reinforcement Learning (DRL) approaches. Here, the Q-values are parametrized using a neural network. Some of the recent DRL approaches including the deep Q-network (DQN) algorithm \cite{dqn}, Double DQN \cite{ddqn}, Dueling network architecture \cite{dueln} and Actor-Critic Methods \cite{ac2016, ac2018}. 

DRL has been used successfully in applications including playing games (such as ATARI). Recently, there has been an increased interest on using DRL for solving computer network problems such as resource allocation, resource management, etc.
However, these are computationally complex and very sensitive to variations in model parameters. There is still substantial amount of work to be done in understanding how these can be applied to large-scale, real-life problems in computer networking.

\begin{figure}
    \centering
    \includegraphics[width=0.9\textwidth]{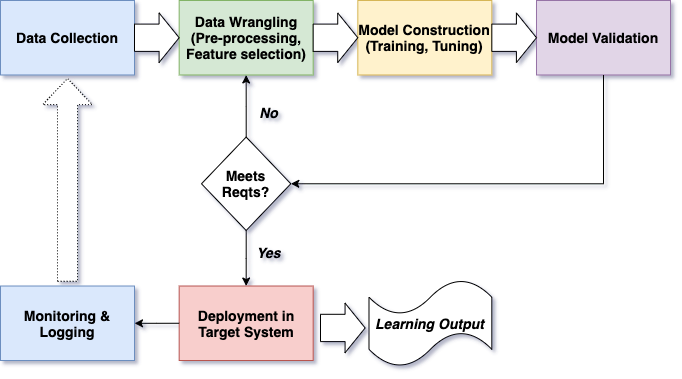}
    \caption{Block diagram of a generic ML-based system and important activities.}
    \label{fig:mls}
\end{figure}

\subsection{ML System}

Fig.~\ref{fig:mls} presents the important activities that are part of an ML-based system. The preliminary step is the determination of the problem statement and identification of the data sources. The first step is collecting the required data sets from an existing system. For instance, in networking systems, this data can be collected from system log files, from network management systems, packet data capture packages such as \textit{tcpdump}, or from other available system interfaces. After collecting the data, the next step is ``data wrangling''. This refers to various data pre-processing activities such as representation and feature selection.

Modeling is the next step in the process. In this step, the decisions regarding use of supervised / unsupervised / semi-supervised choices are made based on the data and the learning tasks or objectives. Subsequently, the specific set of ML techniques to be applied on the data for training  are determined. The baselines are comparison are also identified; these could be based on naive algorithms, heuristics or human expertise. This is followed by splitting the sample data into training and testing datasets. This step also involves activities including feature scaling, regularization and hyperparameter tuning.

In the model validation step, the trained models are tested against the test data and possibly other datasets. This step can also involve cross-validation, where the 
input data is split in $k$ distinct subsets. Next, the first subset is used for testing, with the ML model  trained on the remaining $k-1$ subsets. This process is repeated $k$ times by having a different subset for training in each phase. These different models are then compared in terms of relevant performance metrics, such as precision, recall and accuracy, to select one ML model for further deployment. 

The final trained model is then deployed in the target system. The system is routinely monitored and key performance indicators are logged. This information is then fed back to the collected data, for further refinement of the ML training process. Thus, as time progresses, the system ML models are modified and refined based on prior data sets and live data obtained from the system.

\section{Architectural frameworks and standards}

\label{sec:stds}

The emerging importance of using machine learning techniques for computer networks, has led to  standardization efforts by different industry and standards bodies. This has provided architectural frameworks that incorporate machine learning  in network elements and in the control, management and knowledge plane software entities. Some of these are discussed below.

\subsection{ETSI Experiential Networked Intelligence (ENI)}

The European Telecommunications Standards Institute started the process of creating an architectural framework for incorporating AI/ML in Computer Networks. This has led to the development of Experiential Networked Intelligence (ENI) framework \cite{eni,ENI:CommMag2018,ENI:CommMag2020}. 

The ENI entity provides recommendations to an "Assisted System" (AS) for realizing intelligent network operation and management. This is achieved by applying suitable AI/ML techniques on  observations and state information data collected and provided by the AS. Three different classes of ASs are defined, in decreasing order of capabilities: (i) Class 3: AS with AI functionality and an operational control loop;  (ii) Class 2: AS with AI functionality but without an operational control loop; and (iii) Class 1: AS without any AI functionality. The ENI interacts with the AS using possibly an Application Programming Interface (API) broker that performs necessary translations between the existing AS's API and the ENI's API. 

\begin{figure}
    \centering
 \subfloat[Class 2 Assisted System in Recommender Mode]{
    \includegraphics[width=3in]{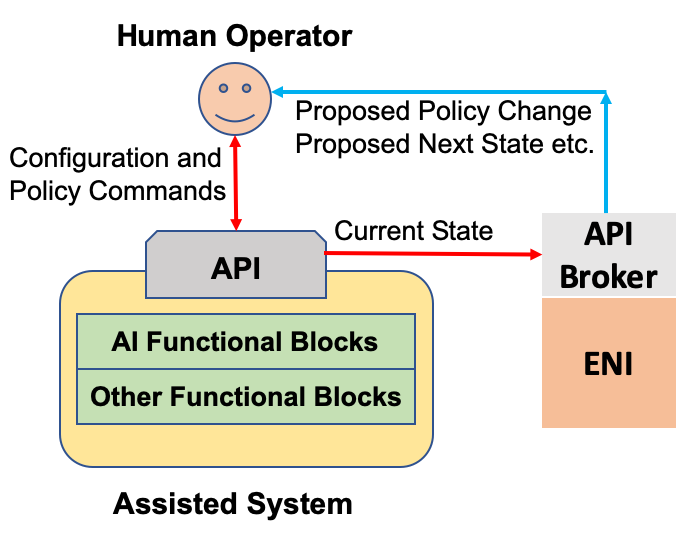}}
    \hfill
    \subfloat[Class 3 Assisted System in Management Mode]{
    \includegraphics[width=3in]{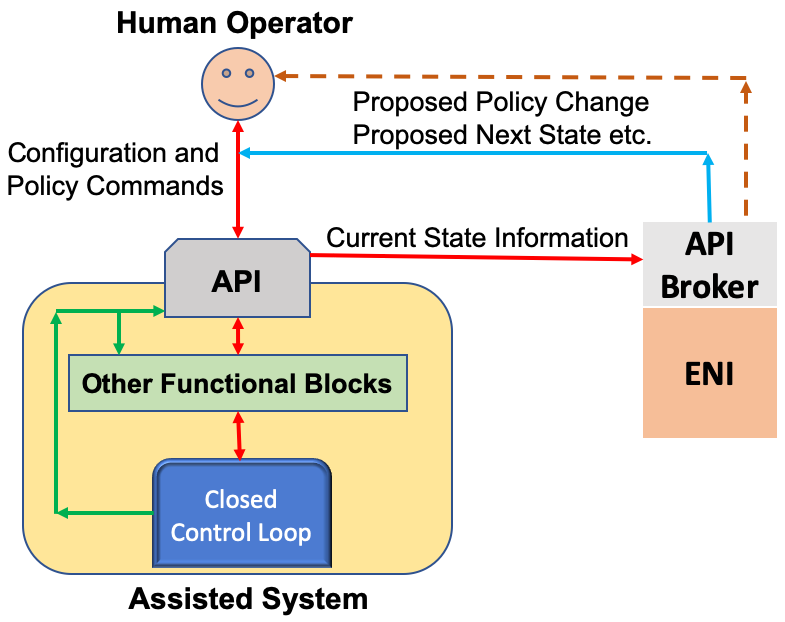}}
    \caption{Two example ENI-AS interaction scenarios, based on \cite{ENI:doc005}.}
    \label{fig:enimodels}
\end{figure}

Based on applying network analytics on the measured data and current system state obtained from the AS, the ENI provides recommendations, which can be in the form of proposed policy changes, proposed next state, and so on. The ENI can operate in: (i) "Recommender" mode, where it provides advice on policy changes to the manual operator of the AS, who maintains control over any changes to the configuration and operation of the AS; or (ii) "Management" mode, where the ENI can provide direct policy commands to the AS, for direct implementation without operator intervention. 

Fig.~\ref{fig:enimodels} presents two possible combinations of AS-ENI interactions. Figs.~\ref{fig:enimodels}(a)  presents a Class 2 system operating in \textit{Recommender} Mode.
Here, the ENI makes recommendations to the human operator who makes the decision about translating the recommendations to specific configuration and policy commands for the AS.
Figs.~\ref{fig:enimodels}(b)  presents a a Class 3 system operating in \textit{Management} Mode.  Here, the ENI makes some recommendations to the human operator and some directly to the API for making necessary changes in the AS. Note that the AS is also enabled with a closed loop AI-based control that uses AI/ML algorithms for adaptively changing system policies and configurations.

\begin{figure}
    \centering
    \includegraphics[width=0.85\textwidth]{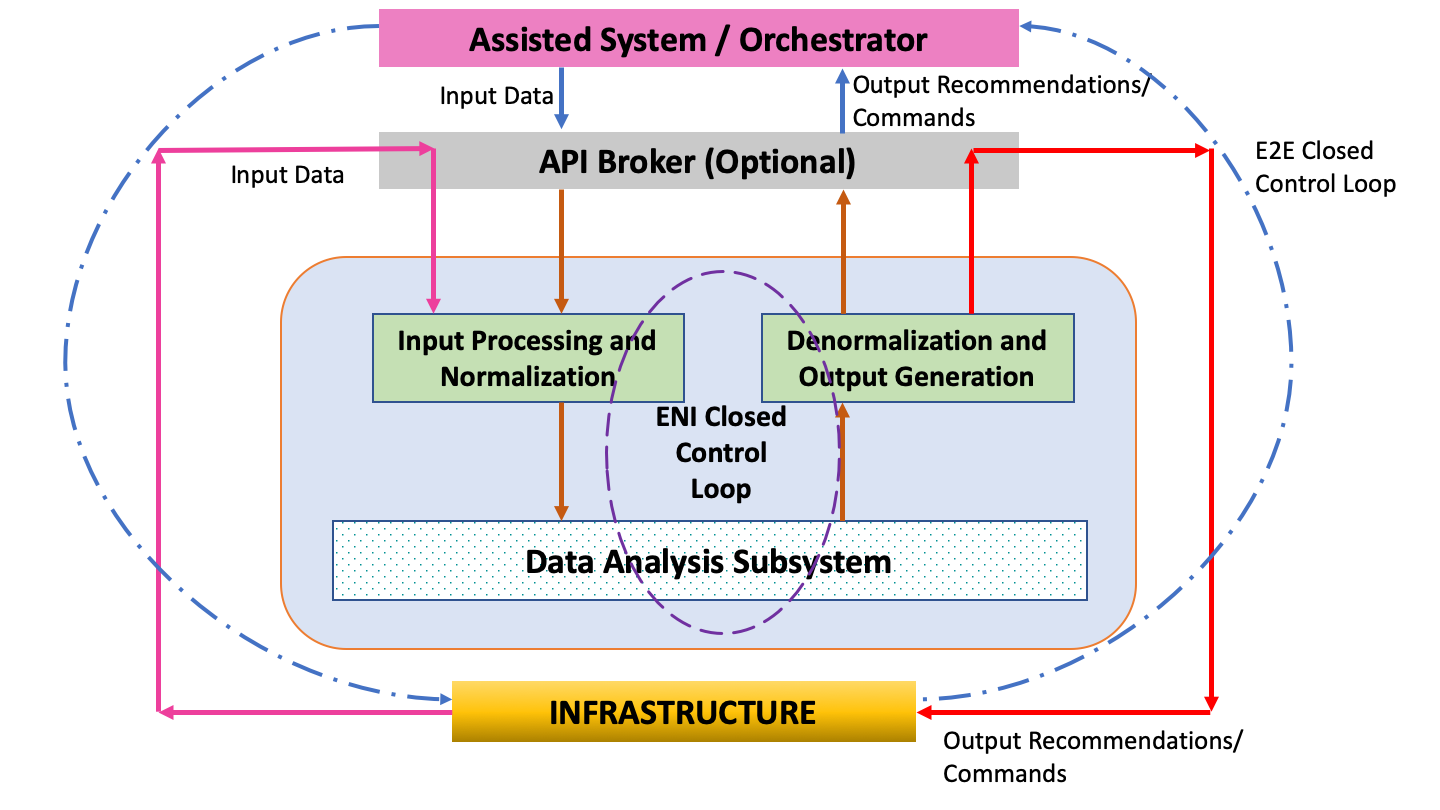}
    \caption{ENI framework depicting ENI-AS interactions, based on \cite{ENI:doc005}, including the End-to-End (E2E) closed control loop between the infrastructure and the assisted system. }
    \label{fig:enistr}
\end{figure}

The internals of the ENI structure are shown in Fig.\ref{fig:enistr}. There, the Assisted System provides input data to the API broker that translates the data into the appropriate ENI format. The input data is further processed and normalized before being fed to the AI/ML network analytics engine. This engine is designed to be context-aware, knowledge-managed and situation-aware.
Internally, the ENI has several functional blocks including those for: (i) Knowledge Representation and Management; (ii) Situational Awareness of the system being analyzed, (iii) Policy-Based Management,  and (iv) Cognition Framework, which actually does all the AI/ML related data analytics and provides the network intelligence capabilities.
The output of the engine in the form of recommendations to the AS, as described earlier, are then fed back to the AS after applying data denormalization and ENI-to-AS data format translation.  

The ETSI ENI group has also identified several possible 5G networks use cases including Service assurance in 5G networks supporting different vertical industries, Intelligent Transport Network Slicing System in 5G, policy-based network slicing for 5G Security, and Intelligent Fronthaul Management and Orchestration. Several proof-of-concept implementation studies related to these use cases have also been presented by various academic and industry groups \cite{eni:poc}.

\begin{figure}
    \centering
    \includegraphics[width=0.9\textwidth]{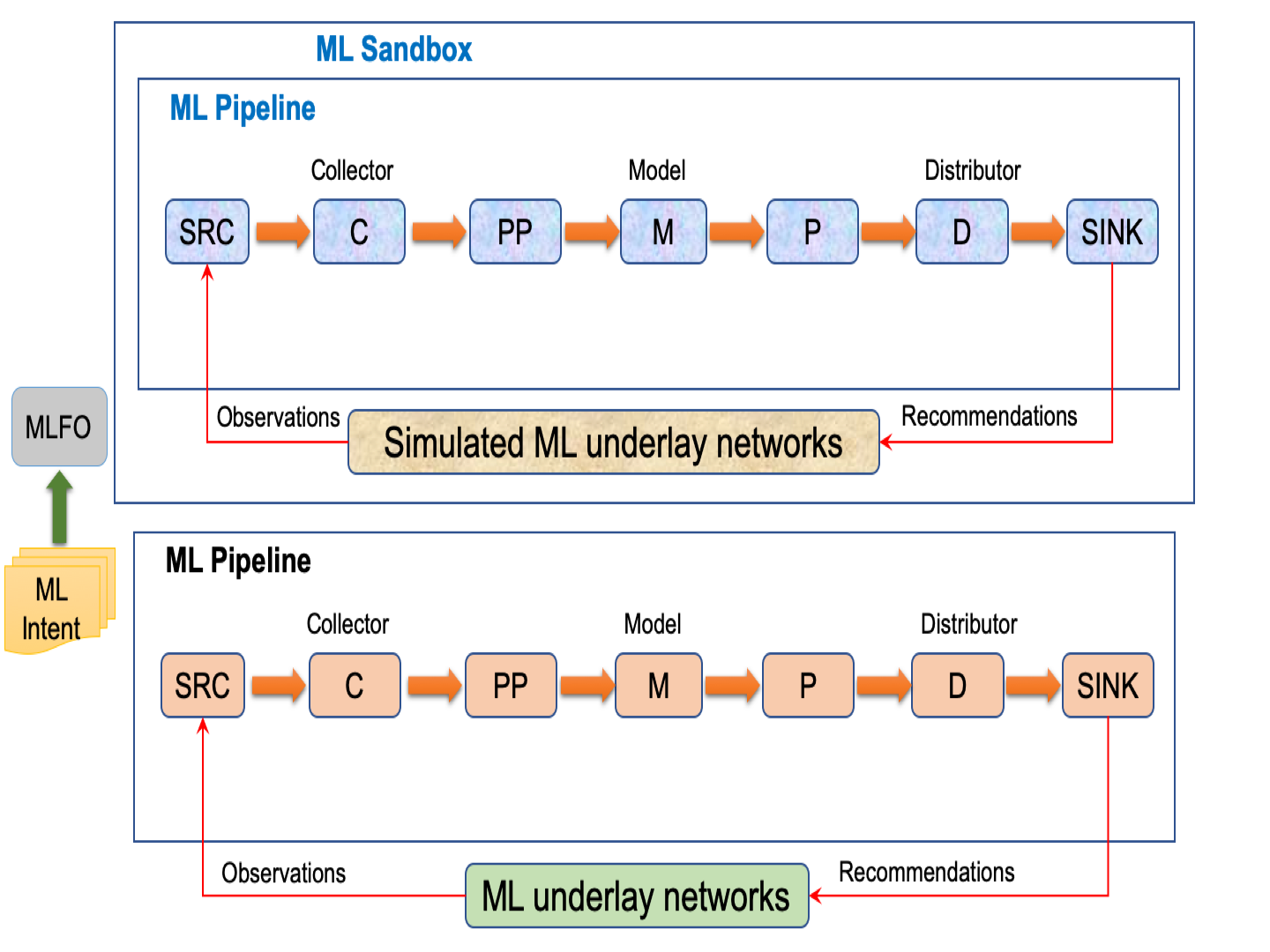}
    \caption{ITU-T Y.3172 machine learning framework, adapted from \cite{itut:y3172}.}
    \label{fig:y3172}
\end{figure}

\subsection{ITU-T Y.3172}

The ITU-T Y.3172 recommendation describes an architectural framework for using machine learning in next-generation networks \cite{itut:y3172}. 
Fig.~\ref{fig:y3172} presents the high-level concepts defined in the Y.3172 framework.
In this framework, the network functionality is realized using a set of network functions (NFs) that can be implemented in dedicated hardware or in virtual network functions (VNFs). The framework incorporates a machine learning pipeline that consists of a set of logical nodes that implement specific AI/ML functionalities. 

The system consists of the underlying network infrastructure, composed of a set of ML underlay networks. Each underlay network is realized using a set of network functions (NFs). The data collected from the underlay networks is fed to the ML pipeline system; similarly, the decisions from the ML pipeline is fed to the underlay networks for making necessary policy and configuration changes in the network elements. The pipeline is made up of a set of nodes, where each node has a specific functionality. Network data is injected into the ML pipeline through the source node (SRC), which obtains data from the underlay network(s). Multiple SRC nodes feed into a data collector node (C), which is followed the pre-processor node (PP). PP performs the necessary data processing including cleaning and aggregation of data. The output from PP is provided a machine learning model (M), which can be supervised or unsupervised as described earlier. The output of the ML model is then fed to the policy node (P), which translates the output from M into actionable policies that will be implemented in the underlying network. These policies are communicated to the underlay network using the distributor and finally to the corresponding sink nodes (e.g. a mobile node, a switch, a router, etc.) that make necessary configuration or policy changes.

The framework also includes the provision to have an ML sandbox with  simulated underlay networks. This enables the system to try out the effects of policy changes obtained from the ML model outputs on the simulated network, before deploying it on the real network. The sandbox ML is also updated in a dynamic manner from the observed system behavior in the real underlay network. This enables the models to be updated on a regular basis with crucial network-level feedback on various key performance indicators (KPIs). 

The network administrator specifies the desired network behaviour using so-called ``ML intents'', which are used by a Machine learning function orchestrator (MLFO) to coordinate amongst the ML pipelines based on the ML intent or current network state. 

The recommendation also provides some example realizations of the above framework in different scenarios, such as  an IMT-2020 network. The 3GPP standard also describes a framework for incorporating network analytics functionality in 5G networks \cite{nwdaf}, but is not presented here due to lack of space.

In addition to standards activities, there is significant industry activity and product development in  designing practical systems that can effectively utilize AI/ML techniques to improve computer networks, as reported in several position papers and white papers \cite{Arista,Ciena,Cisco,Ericsson,ExtremeNets,Juniper,Vmware}.
These products and system tools deal with various perspectives such as user performance, network security and  resource utilization.

\section{Software tools/packages and networking-related data sets}
\label{sec:mlsw}
This section presents some of the commonly used software tools/packages for implementing and datasets that are available for training purposes.

\subsection{ML and RL Software Packages}
There are several public domain packages / tools available for running many ML/DL/RL algorithms. Some of these packages can be run on local machines by downloading the relevant packages or run on cloud instances using technologies such as Google Colab \cite{Colab} and Jupyter notebooks \cite{Jupyter}, using browser interfaces. The 
Google Colab environment also enables users to directly run on GPU/TPU servers located in the cloud (\url{https://cloud.google.com/tpu}), during the development and testing phase. Developers can subsequently run their production applications on the different cloud service providers, including Amazon's AWS, Google Cloud, and Microsoft Azure. 

The ML software packages eliminate the need for network designers and administrators to develop machine learning systems, and enable them to deploy ML-based applications in the network, with minimal programming efforts. Some of the popular ML packages are:

\begin{itemize}
    \item Keras \cite{keras}: Keras is a widely-used deep-learning framework that is built above TensorFlow 2.0 and can be run on Graphical Processing Unit (GPU) clusters or Tensor Processing Units (TPU).
    
    Tensorflow, primarily developed by Google, is an open-source software library for machine learning. It can be used across a range of tasks with specific focus on training and inference of deep neural networks.
    
    \item PyTorch \cite{pytorch}: PyTorch is used to write models for applications such as computer vision and natural language processing. It was primarily developed by Facebook. It provides the advantages of processing in GPUs and automatic differentiation capabilities for deep learning models.
        
    \item Scikit-learn \cite{scikit}: Scikit features various classification, regression and clustering algorithms including support vector machines, random forests, gradient boosting and K-means algorithms. It is designed to inter-operate with the Python numerical and scientific libraries such as NumPy and SciPy.
    
    \item MatLab \cite{matlab}: The MATLAB package provides implementations for several commonly used machine learning algorithms, ready to use out-of-the-box.
   
\end{itemize}

Compared to traditional machine learning algorithms that are used for classification, clustering and regression, there are relatively fewer reinforcement learning packages, and even these do not seem to be well-documented and easy to use, as of this publication date. The primary reason for this could be the fact that RL techniques are still mainly used for research and less for production applications. A few public-domain RL packages are described below:
 
\begin{itemize}
    
    \item OpenAI Gym \cite{gym}: Gym provides the development environment for implementing various reinforcement learning algorithms.

    KerasRL (\url{https://github.com/keras-rl/keras-rl}; Nov. 2019) and KerasRL2 \\(https://github.com/wau/keras-rl2; Mar. 2021) are RL and DRL implementations based on the OpenAI Gym platform. They support many algorithms such as Deep Q-Learning (DQN), Dueling network DQN (Dueling DQN) and Proximal Policy Optimization Algorithms (PPO). 
    
    \item Stable Baselines \cite{stablebl}: This is based on the earlier OpenAI Baselines RL package, and supports several RL algorithms. It also provides the  RL Baselines Zoo \cite{rlblzoo}, a collection of several trained RL Agents that can be used and modified by application developers. 
    
    \item TF Agents \cite{tensorflow}: This is based on Tensorflow and provides agents for RL as well as Contextual Bandits. 
 
\end{itemize}

\subsection{Public-domain Networking Datasets}

One of the significant challenges in realizing ML-based systems is the lack of suitable datasets for both training and benchmarking purposes. Domains such as image processing, computer vision, and natural language processing have several public-domain datasets that have been developed over the years. These datasets enable the application developers to assess the performance of new ML-based schemes. However, this is not the case with networking applications, where there is a substantial need for common datasets. The common techniques to obtain data sets are described below.

\noindent\textbf{Synthetic Generation:} Here, data can be generated using known distributions such as Poisson, Gaussian, etc. with suitable addition of random noise. However, they will not effectively capture real network behavior. For instance, teletraffic tends to exhibit  self-similarity property having long range dependence, which will not be effectively captured by Poisson and other known stochastic processes. 
    
    Hence, methods  to synthesize data from a given data set have been proposed. In \cite{noms2020}, methods for sampling key parameters from live or recorded traffic are described. These samples are used to generate large volumes of synthetic traffic with very similar rate distributions and temporal characteristics.
     
     Generative Adversarial Networks (GANs) \cite{GanGoof2016} can also be used for generating synthetic data sets from a given data set. In \cite{GANFlow2019}, flow-based network data has been generated using GANs, for the purpose of anomaly detection in network traffic. GANs have also been used to share networked time series data between different parties while preserving the underlying traffic characteristics and providing some form of privacy \cite{GANShare2020}.
     
\noindent\textbf{Simulator and Emulator-based Generation: }Discrete-event simulation techniques are used to create detailed models of a networked system. The level of abstraction in such models can be fairly low, by considering abstracted network events such as packet arrivals and departures, packet processing and packet transmissions. Alternatively, the simulation model can be quite detailed by including all details of packet headers, processing, arrivals, departures, transmissions, channel models and so on. Public-domain software such as \textit{ns3} (\url{nsnam.org}) can be used to construct such detailed models of network behavior, where all layers of the network protocol stack are implemented. Such models can be used to generate datasets for machine learning techniques; however, the packet generation process, event generation process (example, malicious behavior of hosts, fault occurrence) are often based on known distributions with some random noise. Thus, these models may not fully generate realistic network traffic, but can be quite accurate in representing networking traffic.
     
    Emulator-based models are also possible to study the performance and behavior of networked systems. A commonly used emulator is \textit{mininet} (\url{mininet.org}),
    that uses the underlying kernel-based network protocol implementation to represent realistic hosts, switches, routers and other network components. The level of protocol implementation is the same as that of a real network. Such emulated network models can also be used to generate network  data sets, but face the same challenges as that of simulation-based models explained above.
    
\noindent\textbf{Measurements and Trace-based Generation:} In this approach, network  traffic is measured and recorded using suitable log files. The collected data is often sanitized and anonymized for privacy reasons, before being released for public use. This provides the most realistic network traffic data. However, there are not many network traffic traces easily available due to reluctance in sharing internal network data and also due to the large data processing requirements when dealing with traffic collected over multi-Gbps and Tbps links over extended periods of time. Some of the publicly available measurements and trace-based networking-related datasets are listed below:

\begin{itemize}
   
    \item Computer Network Intrusion Detection data from KDD Cup 1999:
\url{https://www.kdd.org/kdd-cup/view/kdd-cup-1999}

\item DARPA Intrusion Detection Evaluation Dataset: \url{https://www.ll.mit.edu/r-d/datasets/1999-darpa-intrusion-detection-evaluation-dataset}
    
    \item CRAWDAD datasets: \url{https://crawdad.org/dartmouth/campus/20090909/}, \\ \url{https://crawdad.org/eurecom/elasticmon5G2019/20190828/index.html}
    
    \item iPlane traceroute Dataset, containing Internet topology and performance data: \url{https://web.eecs.umich.edu/~harshavm/iplane/}
    
    \item RIPE Network Coordination Centre Data Repository: \url{https://labs.ripe.net/datarepository/}
    
    \item Center for Applied Internet Data Analysis (CAIDA) Data: 
\url{https://www.caida.org/data/overview/}, \url{https://www.caida.org/data/external/}
    
  \item UC Irvine Machine Learning Repository: \url{https://archive.ics.uci.edu/ml/datasets/Kitsune+Network+Attack+Dataset},
    \url{https://archive.ics.uci.edu/ml/datasets/detection_of_IoT_botnet_attacks_N_BaIoT}
    
    \item City of Milano: Traces are available for telecommunications data collected in the City of Milano, by the Telecom Italia cellular network, available at \url{https://dataverse.harvard.edu/dataset.xhtml?persistentId=doi:10.7910/DVN/EGZHFV}, \url{https://dataverse.harvard.edu/dataset.xhtml?persistentId=doi:10.7910/DVN/JZMTBJ}.
    
    \item MAWI Working Group Traffic Archive: \url{http://mawi.wide.ad.jp/mawi}: samplepoint-F monitors the 1Gbps transit link in Tokyo to an upstream provider. The traffic is between WIDE and its non-peer ASes. samplepoint-G monitors the 10Gbps link to DIX-IE, an experimental IX in Tokyo operated by WIDE. 
    
    \item Kaggle datasets: This provides several datasets at 
    \url{https://www.kaggle.com/datasets}, an example of which is \\ \url{https://www.kaggle.com/ic17b035/netstatistics-from-rtr-periode-2017-to-2019}
    
    \item Commercial 5G Cellular network performance data in Minneapolis, MN; Chicago, IL; Atlanta, GA, collected in 2020 and     available at \url{https://fivegophers.umn.edu/www20/}.
    
    \item NYU Metropolitan Mobile Bandwidth Trace, a.k.a. NYU-METS, is a LTE mobile bandwidth dataset that was measured in the New York City metropolitan area, available at \url{https://github.com/NYU-METS/Main}.
    
  \end{itemize}  
    
There are several other publicly disseminated datasets available at kaggle, crawdad, UCI and other repositories.

\section{Recent Applications of ML techniques in Networks}
\label{sec:uses}

This section presents a sample set of studies, that demonstrate the use of ML techniques in computer networking.

\subsection{Packet Classification in Routers}

Packet classification is one of the fundamental tasks performed by a
router. It helps with detecting network attacks, traffic engineering,
quality of service scheduling, etc. Each router has a set of rules
that is applied to each incoming packet in order to classify it. This
task has been traditionally done either using a specialized hardware
unit called Ternary Content-Addressable Memory (TCAM), or using
software approach with specialized  data structures such as a
retrieval tree (Trie) or decision trees.  TCAMs can execute all the
rules in parallel and provide the results in constant time. However,
TCAMs are complex and have high cost and power consumption and might
not scale to large number of rules and table sizes. Software
approaches are scalable but take longer time as compared to the
hardware approach. Creating an efficient Decision Tree is difficult as
it depends on a hand-tuned heuristics and these heuristics cannot
explicitly optimize for a given objective like the tree depth. Hence,
creating an optimal decision tree is necessary, considering
high-scalability and fast-computation requirements.  

The work presented in \cite{Ref:NueralPacketClassification} shows how
ML techniques can be applied to the classification problem. This paper
proposes \textit{NeuroCuts}, a Deep Reinforcement learning based
classification algorithm. Based on a rule set and an objective
function (i.e., classification time, memory footprint or both),
NeuroCuts learns to build an optimal decision tree.

As an RL system, NeuroCuts uses the set of Rules and the current
decision tree as the environment. The agent is a DNN model, which aims
to select the best cut or partition action for building the tree. This
cut divides a node along a dimension (i.e., one of SrcIP, DstIP,
SrcPort, DstPort, and Protocol) into a number of sub-ranges, creating
many child nodes. The partition action divides rules into disjoint
subsets, creating new child nodes for each subsets.  A rollout is a
sequence of actions that build a decision tree with the associated
rewards by using a given policy. The agent starts with an initial
state i.e. a single node. At each step, agent performs an action and
evaluates it with multiple rollouts and updates it based on the rewards
received. Based on this action, environment changes the state and the
process is repeated. The goal is to maximize the total reward received
by the agent.

Benchmarks \cite{Ref:Classbench} is used to generate packets
classifiers with different characteristics and sizes. NeuroCuts
converges to an optimal solution within just a few hundred rollouts,
where each rollout has up to 15000 actions.  Neurocut is also compared
with other approaches such as HiCuts \cite{Ref:HiCuts}, HyperCuts
\cite{Ref:HyperCuts}, EffiCuts \cite{Ref:EffiCuts} and CutSplit
\cite{Ref:CutSplit}, in terms of classification time (tree depth) and
memory footprint (bytes per rule).  The best time-optimized NeuroCuts
trees provide 20\%, 38\%, 52\%, and 56\% median improvement over
HiCuts, HyperCuts, EffiCuts, and CutSplit
respectively. Space-optimized NeuroCuts trees show 40\% median
improvement over EffiCuts, significantly better than HiCuts and
HyperCuts. It cannot outperform CutSplit in memory footprint.

\subsection{User Traffic Prediction System in Cellular Networks}

%https://doi.org/10.1109/TMC.2020.3001225

%This section summarizes the work done on traffic %prediction in cellular networks, presented in %\cite{STEP2020}.

To meet the increasing cellular traffic, small cells (micro,femto,
pico) have been widely deployed to increase the network capacity. But
a moving user may experience frequent handoffs and as per the current
handoff procedure (LTE standard Rel-8) the decisions are made based on
real-time signal strength measurements from neighboring BSs which
leads to frequent handoffs between the BSs. As a result, there will be
an increase in call dropping probability (CDP) due to the latency
incurred during the handoffs signaling between the BSs. This is going
to affect the reliability of conventional handoff procedures in
handling high mobility low latency services in 5G and beyond. To
overcome this issue proactive cell selection and reservation schemes
have been proposed in the literature to assist handoffs and reduce
CDP, which depend heavily on accurate prediction of users’ traffic and
mobility.

There have been several research studies on large-scale traffic
prediction i.e., forecasting the traffic load or number of subscribers
for a city, an urban area, or a cell, based on data aggregated by the
network administrators. Such a prediction is too coarse to be applied
to assist in handoffs or to predict individual cellular traffic
usage. In \cite{STEP2020}, a spatio-temporal fine granular traffic
prediction system for cellular networks has been proposed.  The
objective is to assist the network operator to reserve transmission
resource blocks (RBs) for individual users in a radio access network
at basestations that the users might access soon, as suggested by a
per-user traffic and mobility prediction algorithm. Generally, traffic
forecasting is modeled as a time series forecasting problem. However,
the analytical models used do not adequately capture the complex and
nonlinear hidden spatio-temporal dependency in wireless traffic
data. To overcome this issue, a technique called GCGRN, based on
combining graph convolution network (GCN) \cite{GCNSurvey2019} and
gated recurrent units (GRU), has been developed to capture
Spatio-temporal usage patterns of the individual user traffic. GCNs
have been used since the user traffic is represented as
graph-structured data and also since they are computationally more
efficient than graph neural networks (GNNs) \cite{GNNSurvey2021}. The
GRU, which can be considered similar to the Long Short-Term Memory
(LSTM) technique, is used to model temporal dependency \cite{GRU2014}.

The training dataset comprises traffic statistics and geolocations
data collected from 10 volunteers over one month period (from
03/01/2019 to 04/01/2019) through a specialized data collection
application. Each entry consists of the phone’s time instance, the
connected BS (Base Station) ID, GPS coordinate, overall cellular
uplink/downlink data rate, and cellular uplink/downlink data rate for
each online app being sampled at one-second frequency. Thus, the
resulting dataset contains 106 data samples involving 1564 BSs and 12
online apps. This was further augmented with synthetic data using the
Slicing Window technique to scale data to 150 users to capture dynamic
service usage patterns.  The results show that GCGRN prediction
accuracy is better than ARIMA, LSTM, and GNN in terms of Mean Absolute
Relative Error (MARE) and Root Mean Squared Error (RMSE). It is also
shown that GCGRN BS prediction accuracy during handoffs is better than
that obtained by using Graph Neural Networks (GNN).

Online training produces a more accurate prediction over offline
training as it consistently updates the model to accommodate
unforeseen traffic in a timely manner. In addition, network-level
simulations using the \textit{ns3} package, show that GCGRN based
resource reservation effectively reduces session dropping ratio during
handoffs in cellular networks.

\subsection{Wi-Fi Intrusion Detection System}

Intrusion detection systems (IDS) are used in networks as an important
security tool. These systems monitor network traffic regularly and
identify any anomalous traffic that could indicate a malicious attack
on the network. IDS systems have to handle significant amount of
traffic meta-data and operate in near real-time in order to identify
intrusions. IDS systems tend to use sets of rules and machine learning
models to identify anomalies.

The work presented in \cite{satam2020wids} presents a Wi-Fi Intrusion
Detection System (WIDS) that performs anomaly detection using ML
models. In the last couple of decades, Wi-Fi has undergone a lot of
improvements in speed, reliability and also in security aspects. Even
though lot of enhancements in Wi-Fi security standards such as IEEE
802.11i, IEEE 802.11w, Wi-Fi Protected Access 3 (WPA3), Open Wireless
Encryption (OWE) are proposed, Wi-Fi networks are still vulnerable to
many attacks like de-authentication attack, man-in-the-middle (MITM)
attack, Denial of Service (DoS) attack etc. Hence, we need a WIDS
system to detect these Wi-Fi intrusion with low false positives.
	
Wi-Fi traffic is modeled as normal or malicious using $n$-grams (a
sequence of $n$ words), observation flows data structures and ML
classification techniques. In this work, anomaly based WIDS is
implemented by monitoring the Wi-Fi protocol state transition machine.
Features are extracted from raw Wi-Fi frames and converted into
observation-wireless (OWF) flows based on $n$-grams. This is used to
calculate the probability of flow, total frames in the flow, total
number of new $n$-grams in the flow, management ratio (Ratio of number
of management frames to total frames in the flow), control ratio, data
ratio. These are then used as features for building ML models to
classify normal and abnormal behavior, using a binary classifier. 
	
%% Let $M$ be the memory, where normal models are extracted from normal
%% events $N^R$, $f$ be the normal behavior characterization function,
%% which is defined as $f:U^R \times M \rightarrow [0,1]$, where $U^R =
%% N^R \cup A^R$ and a detector (binary classifier) defined as $D=(f,M)$,
%% which classify the samples as normal (0) or abnormal (1).
	
Wi-Fi traffic data was collected using 6-7 Wi-Fi access points (AP)
and around 15-200 users to obtain a normal training set for 9,14 and
38 days. In addition, the abnormal dataset was collected using libpcap
scripts using a Linux-server based access point and known attacks,
listed below.  The public Aegean Wi-Fi Intrusion Dataset (AWID)
dataset was also used (\url{https://icsdweb.aegean.gr/awid}). Using
the $n$-gram size analysis, it was observed that the data collected
for 12-15 days is adequate to obtain a nearly complete understanding
of Wi-Fi normal behavior.

Different kinds of Wi-Fi attacks such as Wi-Fi availability attacks
(De-authentication Attack, Disassociation Attack, etc) and Wi-Fi
encryption attacks (Dictionary attack. Fragmentation attack, etc) are
detected with high accuracy and less false alarm. A performance study
of ML classification algorithms such as Isolation Forest, C4.5, Random
Forest, AdaBoost and Decision Table with respect to different Wi-Fi
attacks was done; the study showed that the Isolation Forest performed
better overall.  Further, a new minimal footprint attack called
minimal de-authentication attack was presented. It is shown that the
WIDS system with C4.5 ML model detects this attack as well. 
	
In the experiments, only the attacks targeting the availability of
Wi-Fi networks are covered, not the encryption. Recent attacks in
Wi-Fi such as the Krack attack and Dragonblood attack on WPA2 and WPA3
protocols are not covered in this article, and can be studied as part
of future work.

\subsection{Learning based Routing in Circuit-Switched Networks}

In circuit-switched networks, connections are established between a
given source and destination at the time of creating the connection,
on a specific path. All the data on the connection subsequently flows
on this path. Routing algorithms are used to determine the suitable
path for a given connection request. The main objectives are to admit
the maximum number of connections, or equivalently, minimize the
connection dropping probability and to minimize the resource
utilization on the network links. The routing algorithms use static
paths as in shortest path routing and fixed-alternate path routing
schemes or dynamic paths as in adaptive routing schemes. Among
dynamic, the least loaded (LL) routing algorithm that considers the
links' utilization status at the time of connection established is
widely used. The LL algorithm selects the path with the least load in
the network, but has been shown to exhibit inefficiencies under
certain network conditions.

In \cite{NBLLRoute2019}, a supervised naive Bayes classifier approach
is used as part of the routing algorithm. It uses historical network
state information to predict the future connection blocking
probability and selects paths that will minimize this value, each time
a connection is established. When a new connection request in
received, the algorithm selects the path based on the Bayes classifier
(which uses prior network state snapshots) and also records the
current network state information. The classifier is updated
periodically based on the past information and current network status.
This algorithm is incorporated in a SDN-based network controller and
has also been implemented using a parallel computing framework in
order to speed up the learning and decision-making process.

The algorithm has been studied for two network topologies: (i)
14-node, 21-link NSFNET network and (2) 21-node, 25-link ARPA-2
network. Compared to existing LL algorithm, the proposed approach is
seen to outperform in terms of reduced connection blocking probability
and in terms of lower network capacity consumption.

\subsection{Optimizing Resource Provisioning in 5G Network Slicing
  With Capacity Forecasting}

Network Slicing enables the customization of virtual networks, offered
to various mobile services over the same physical infrastructure. It
is a key enabler of heterogeneous mobile applications in 5G and
beyond-5G mobile networks.  In such networks, multiple network slices
share the infrastructure resources that are orchestrated according to
the varying traffic demands of individual slices. The slicing concept
is realized using network function virtualization (NFV), where the 5G
Core and RAN software are implemented as a set of independent and
interacting virtual network functions (VNF). Each VNF is realized
using commodity computing hardware (e.g. servers) running a
virtualized environment consisting of virtual machines or containers.
Thus, each slice requires a set of VNF resources to deployed in a
shared computing environment, such as a data center.

Current resource orchestration approaches are often reactive in
practice relying on fault-tolerance and self-healing. However, such
solutions can lead to service degradation.  Effectively sharing
resources among the slices while managing the trade-off between system
utilization and management costs is challenging and mandates a
proactive approach. Hence, proactive resource management solutions for
network slicing have been popular in recent years and most of them
either use redundancy or traffic prediction modules to accommodate
varying network traffic conditions.

Mobile network traffic forecasting is an extensively studied
subject. As described in \cite{DeepCog}, these solutions often focus
on predicting the expected traffic volumes in the near future.  Such
an approach is not suitable for network slicing where resource
allocation can have a profound impact on monetary costs.  Using
traffic predictors or redundancy can lead to either overprovisioning
or Service Level Agreements (SLA) violations. The \textit{DeepCog}
architecture, proposed in \cite{DeepCog}, is a deep learning based
cost-aware capacity forecasting solution aimed at solving these
challenges to enable efficient dynamic resource management in 5G
networks. It differs from traditional traffic forecasters by taking
into account the monetary costs related to underprovisioning or
overprovisioning of resources.

\textit{DeepCog} is designed to be a slice aware resource management
framework working at the datacenter level. Its capacity forecasts are
made for each slice and represent the amount of traffic expected at
the datacenter for that slice. The framework collects the slice
traffic observed across $N$ basestations represented as \begin{math}
  \delta_s(t) = \{ \delta_s^1(t),...,\delta_s^N(t) \} \end{math}. This
information is then aggregated for $M$ datacenters to define the slice
demand at datacenter $j$ for slice $s$ as \begin{math}d_s^j(t) =
  \sum_{i|f(i)=j}\delta_s^i(t)\end{math}. Using this information,
DeepCog predicts a constant capacity $c_s^j(t)$ of slice $s$ for the
time period $(t,T_h)$ at datacenter $j$.  Given the limitations of NFV
technologies reconfiguration capabilities, infrastructure providers
would have to maintain constant capacity of the reconfiguration time
period represented as $T_h$. The deep learning framework used to
predict these capacities is described below using three major
components.

DeepCog is realized using a deep neural network (DNN) architecture,
with a 3D-CNN (Convolutional Neural Network) acting as the encoder and
Multi-Layer Perceptrons (MLPs) as the decoder. The slice traffic
aggregated from various basestations is transformed to a 3D-tensor,
which in turn is fed to the encoder network. The motivation behind
using a 3D-CNN is to extract spatio-temporal features in the traffic
data.  Given the recent success of 2D-CNNs for image and video
processing, the 3D version of CNNs was used due to the additional time
dimension in network traffic data. he encoder uses three 3D-CNN layers
to process the tensor input where each layer is made up of neurons
running a specific filter.

The idea is to limit the receptive field in the first layer in order
to emphasize strong local correlation. The encoder's output is a
uni-dimensional vector which is then fed to the MLP based decoder. The
decoder consists of four neural layers, including the output layer,
where all of them are fully connected and use ReLU as the activation
function. The output layer though uses a linear activation function as
the capacity forecasts need to be real values. The output layer is
configured to output multiple predictions in parallel each
representing a subset of basestations. The intuition behind using an
MLP as a decoder is its ability to learn complex functions.

Raw input data traffic is collected using various probes offered by
network infrastructure. DeepCog transforms the traffic data into a
3D-tensor and ensure local correlation among nearby elements. Instead
of using geographical location to collocate the traffic data, DeepCog
rather relies on calculating the shape based distance (SBD) between
recorded demands at basestations i and j for all pairs.  SBD between
each pair is computed using historic traffic observed at the
basestation. These SBD values are further used to calculate virtual
bi-dimensional coordinates and mapped to the elements of 3D-tensor by
solving an optimization and an assignment problem.

The Adam optimizer \cite{Adam2014} is used for training the neural network.  The
encoder-decoder are not trained together as one block; encoder is
trained on maximum time granularity, with decoders tailored to each
time granularity $T_h$. This is possible due to the
horizon-independence of the encoder and it also reduces the total
training time significantly. At each SGD iteration, the predictions
are evaluated using a customized loss function defined using a
parameter $\alpha_{sj}$. This represents the ratio between SLA
violation costs and capacity costs for slice $s$ at datacenter $j$.  A
higher value of $\alpha_{sj}$ translates to prioritizing
overprovisioning over system utilization.  More details on the loss
function can be found in the paper.

DeepCog was analyzed using real world measurement data of a large
operator, not disclosed, over an area of 100$km^2$ and 2 million
inhabitants. The data collected was categorized into social networks,
messaging services and video streaming applications and are further
mapped to different datacenter classes: Core, MEC and C-RAN
respectively and they represent the variation in number of
basestations covered presenting a diverse case study. The model was
evaluate over the time horizon ($T_h$) ranging from 5 minutes to 8
hours, with the observation window $T_p$ fixed to 30 minutes. First
the model was evaluated against other existing traffic predictors and
was shown to perform exceedingly well, with DeepCog only incurring 15
to 27\% of the costs incurred by the best predictor across the three
use-cases. The model was also evaluated over various values of
$\alpha$ and it was shown that the parameter achieves the desired
behaviour i.e. higher value of $\alpha$ leading to lower SLA
violations. Further the authors evaluate the impact of the duration of
time horizon on the monetary costs and have observed a quasi-linear
relationship showing that costs do not skyrocket with increased $T_h$.
The dataset is not available, but the source code for DeepCog is
publicly hosted \cite{DeepCogSC}.

This section presented a small sample of application of machine
learning algorithms to solve problems in computer networking. There
are several such studies, many of which are summarized in various
survey articles
\cite{dqn,JISA:Surv18,ShenVTS20,KBN:CCR17,AI:OSN2018,Kafle2018,Bin2019,DataDriven5G2020}.

\section{Conclusions}
\label{sec:concl}
This paper presented an overview of machine learning applications for
computer networks. The paper presented a brief overview of the
important ML tasks and highlights of some of the commonly used ML
algorithms, including those used for supervised, unsupervised and
reinforcement learning. The availability of ready-to-use ML software
packages and cloud computing environments in recent years has
significantly contributed to the adoption of ML algorithms in many
domains including computer networking. The paper summarized some of
the popular ML packages and also data sets related to networking. Case
studies from different recent research activities that successfully
demonstrated the use of ML in computer networking were also
presented. However, this is only the tip of the iceberg, as far as ML
for networks is considered. Future generation network design and
implementation will consider machine learning, analytics and other
aspects as an integral part of the network. This is expected to
greatly improve resource utilization, user satisfaction and also
reduce overall network operational costs due to higher
efficiency. There is room for significant innovation both in the
development of ML techniques that are better designed for handling
computer networking data and for real-time network operations, and in
the development of network architectures, frameworks and mechanisms
that can exploit the advanced machine learning capabilities. \\

%% The author is extremely grateful and
%% indebted to the reviewers who provided valuable suggestions and
%% comments to significantly enhance the quality of this paper.

{\small
\noindent\textbf{Acknowledgments:} The author is thankful to Dr. Mathai Joseph
of ACM India and Dr. Sudip Misra of IIT Kharagpur for suggesting this
topic. The author is also thankful to the following research scholars
and project associates of DAWNLAB in the CSE Department of IIT Madras,
for their various contributions towards this paper: Radhakrishna
Kamath, Saurav Chakraborty, Shivani Saxena, Akhil Polamarasetty, Sai
Kiran Posam and Kavin Thangadurai. The author is grateful to
Dr. Madanagopal Ramachandran and his Machine Learning team of
NMSWorks, Chennai for the several fruitful discussions.
}
\bibliographystyle{IEEEtran}
%\bibliography{MLNets}

\begin{thebibliography}{10}
\providecommand{\url}[1]{#1}
\csname url@samestyle\endcsname
\providecommand{\newblock}{\relax}
\providecommand{\bibinfo}[2]{#2}
\providecommand{\BIBentrySTDinterwordspacing}{\spaceskip=0pt\relax}
\providecommand{\BIBentryALTinterwordstretchfactor}{4}
\providecommand{\BIBentryALTinterwordspacing}{\spaceskip=\fontdimen2\font plus
\BIBentryALTinterwordstretchfactor\fontdimen3\font minus
  \fontdimen4\font\relax}
\providecommand{\BIBforeignlanguage}[2]{{%
\expandafter\ifx\csname l@#1\endcsname\relax
\typeout{** WARNING: IEEEtran.bst: No hyphenation pattern has been}%
\typeout{** loaded for the language `#1'. Using the pattern for}%
\typeout{** the default language instead.}%
\else
\language=\csname l@#1\endcsname
\fi
#2}}
\providecommand{\BIBdecl}{\relax}
\BIBdecl

\bibitem{AIRussell2019}
S.~Russell and P.~Norvig, \emph{{Artificial Intelligence: A Modern Approach}},
  4th~ed.\hskip 1em plus 0.5em minus 0.4em\relax Pearson, 2019.

\bibitem{McKinsey2017}
J.~Bughin, E.~Hazan, S.~Ramaswamy, M.~Chui, T.~Allas, P.~Dahlstrom, N.~Henke,
  and M.~Trench, ``Artificial intelligence: the next digital frontier?''
  \url{https://apo.org.au/node/210501}, McKinsey Global Institute, Tech. Rep.,
  Jun. 2017.

\bibitem{HBR2018}
T.~H. Davenport and R.~Ronanki, ``{Artificial Intelligence for the Real
  World},'' \emph{Harvard Business Review}, vol.~1, pp. 1--10, Jan.-Feb. 2018.

\bibitem{ExplAI2019}
\BIBentryALTinterwordspacing
D.~Gunning and D.~Aha, ``{DARPA’s Explainable Artificial Intelligence (XAI)
  Program},'' \emph{AI Magazine}, vol.~40, no.~2, pp. 44--58, Jun. 2019.
  [Online]. Available:
  \url{https://ojs.aaai.org/index.php/aimagazine/article/view/2850}
\BIBentrySTDinterwordspacing

\bibitem{dqn}
V.~Mnih, K.~Kavukcuoglu, D.~Silver, A.~A. Rusu, J.~Veness, M.~G. Bellemare,
  A.~Graves, M.~Riedmiller, A.~K. Fidjeland, G.~Ostrovski, S.~Petersen,
  C.~Beattie, A.~Sadik, I.~Antonoglou, H.~King, D.~Kumaran, D.~Wierstra,
  S.~Legg, and D.~Hassabis, ``Human-level control through deep reinforcement
  learning,'' \emph{Nature}, vol. 518, no. 7540, pp. 529--533, Feb. 2015.

\bibitem{JISA:Surv18}
R.~Boutaba, M.~A. Salahuddin, N.~Limam, S.~Ayoubi, N.~Shahriar,
  F.~Estrada-Solano, and O.~M. Caicedo, ``A comprehensive survey on machine
  learning for networking: evolution, applications and research
  opportunities,'' \emph{Journal of Internet Services and Applications},
  vol.~9, no.~16, 2018, \url{https://doi.org/10.1186/s13174-018-0087-2}.

\bibitem{Wang2018}
M.~{Wang}, Y.~{Cui}, X.~{Wang}, S.~{Xiao}, and J.~{Jiang}, ``{Machine Learning
  for Networking: Workflow, Advances and Opportunities},'' \emph{IEEE Network},
  vol.~32, no.~2, pp. 92--99, 2018.

\bibitem{ShenVTS20}
X.~Shen, J.~Gao, W.~Wu, K.~Lyu, M.~Li, W.~Zhuang, X.~Li, and J.~Rao,
  ``{AI-Assisted Network-Slicing Based Next-Generation Wireless Networks},''
  \emph{IEEE Open Journal of Vehicular Technology}, vol.~1, pp. 45--66, Feb.
  2020.

\bibitem{KBN:CCR17}
A.~Mestres, A.~Rodriguez-Natal, J.~Carner, P.~Barlet-Ros, E.~Alarc\'{o}n,
  M.~Sol\'{e}, V.~Munt\'{e}s-Mulero, D.~Meyer, S.~Barkai, M.~J. Hibbett,
  G.~Estrada, K.~Ma'ruf, F.~Coras, V.~Ermagan, H.~Latapie, C.~Cassar, J.~Evans,
  F.~Maino, J.~Walrand, and A.~Cabellos, ``Knowledge-defined networking,''
  \emph{SIGCOMM Comput. Commun. Rev.}, vol.~47, no.~3, p. 2–10, Sep. 2017.

\bibitem{AI:OSN2018}
J.~Mata, I.~{de Miguel}, R.~J. Durán, N.~Merayo, S.~K. Singh, A.~Jukan, and
  M.~Chamania, ``{Artificial intelligence (AI) methods in optical networks: A
  comprehensive survey},'' \emph{Optical Switching and Networking}, vol.~28,
  pp. 43--57, 2018.

\bibitem{Kafle2018}
V.~P. {Kafle}, Y.~{Fukushima}, P.~{Martinez-Julia}, and T.~{Miyazawa},
  ``{Consideration On Automation of 5G Network Slicing with Machine
  Learning},'' in \emph{Proc. of ITU Kaleidoscope: Machine Learning for a 5G
  Future (ITU K)}, 2018, pp. 1--8.

\bibitem{Bin2019}
B.~Han and H.~D. Schotten, ``{Machine Learning for Network Slicing Resource
  Management: A Comprehensive Survey},'' \emph{ZTE Communications}, vol.~40,
  no.~2, pp. 27--32, Sep. 2019.

\bibitem{DataDriven5G2020}
B.~{Ma}, W.~{Guo}, and J.~{Zhang}, ``{A Survey of Online Data-Driven Proactive
  5G Network Optimisation Using Machine Learning},'' \emph{IEEE Access},
  vol.~8, pp. 35\,606--35\,637, 2020.

\bibitem{NetAI2020}
\emph{Proceedings of the ACM SIGCOMM Workshop on Network Meets AI \& ML
  (NetAI)}, Aug. 2020.

\bibitem{IEEEAIWS2020}
\BIBentryALTinterwordspacing
``{IEEE Workshop on AI/ML in Networks and Cloud},'' New Brunswick, NJ, USA,
  Feb. 2020. [Online]. Available:
  \url{http://www.ieee-ai.org/docs/AL-ML-Agenda-02-14-2020.pdf}
\BIBentrySTDinterwordspacing

\bibitem{eni}
``{ETSI -- Experiential Networked Intelligence (ENI)},''
  \url{https://www.etsi.org/technologies/experiential-networked-intelligence},
  Mar. 2021.

\bibitem{ENI:CommMag2018}
Y.~{Wang}, R.~{Forbes}, C.~{Cavigioli}, H.~{Wang}, A.~{Gamelas}, A.~{Wade},
  J.~{Strassner}, S.~{Cai}, and S.~{Liu}, ``{Network Management and
  Orchestration Using Artificial Intelligence: Overview of ETSI ENI},''
  \emph{IEEE Communications Standards Magazine}, vol.~2, no.~4, pp. 58--65,
  2018.

\bibitem{ENI:CommMag2020}
Y.~{Wang}, R.~{Forbes}, U.~{Elzur}, J.~{Strassner}, A.~{Gamelas}, H.~{Wang},
  S.~{Liu}, L.~{Pesando}, X.~{Yuan}, and S.~{Cai}, ``{From Design to Practice:
  ETSI ENI Reference Architecture and Instantiation for Network Management and
  Orchestration Using Artificial Intelligence},'' \emph{IEEE Communications
  Standards Magazine}, vol.~4, no.~3, pp. 38--45, 2020.

\bibitem{nwdaf}
``{3GPP Release 17 -- 5G Security Assurance Specification (SCAS); Network Data
  Analytics Function (NWDAF)},''
  \url{https://www.3gpp.org/ftp/Specs/archive/33_series/33.521/}, Feb. 2021.

\bibitem{itut:y3172}
``{ITU-T Y.3172: Architectural framework for machine learning in future
  networks including IMT-2020},''
  \url{https://www.itu.int/rec/T-REC-Y.3172-201906-I/en}, Jun. 2019.

\bibitem{KnowPlaNet2003}
D.~D. Clark, C.~Partridge, J.~C. Ramming, and J.~T. Wroclawski, ``A knowledge
  plane for the internet,'' in \emph{Proc.of ACM SIGCOMM}, 2003, p. 3–10.

\bibitem{ImbClass2009}
Y.~Sun, A.~K.~C. Wong, and M.~S. Kamel, ``{Classification OF Imbalanced Data: A
  Review},'' \emph{International Journal of Pattern Recognition and Artificial
  Intelligence}, vol.~23, no.~04, pp. 687--719, 2009.

\bibitem{drl2018}
V.~Francois-Lavet, P.~Henderson, R.~Islam, M.~G. Bellemare, and J.~Pineau,
  ``{An Introduction to Deep Reinforcement Learning},'' \emph{Foundations and
  Trends in Machine Learning}, vol.~11, no. 3-4, pp. 219--354, Dec. 2018.

\bibitem{CourseraNgML}
A.~Ng, ``{Machine Learning},''
  \url{https://www.coursera.org/learn/machine-learning}, 2021.

\bibitem{xgboost}
T.~Chen and C.~Guestrin, ``{XGBoost: A Scalable Tree Boosting System},'' in
  \emph{Proc.of the 22nd ACM SIGKDD International Conference on Knowledge
  Discovery and Data Mining}, 2016, p. 785–794.

\bibitem{Goodfellow2016}
I.~Goodfellow, Y.~Bengio, and A.~Courville, \emph{{Deep Learning}}.\hskip 1em
  plus 0.5em minus 0.4em\relax MIT Press, 2016,
  \url{http://www.deeplearningbook.org}.

\bibitem{Qlearn}
\BIBentryALTinterwordspacing
C.~J. C.~H. Watkins, ``Learning from delayed rewards,'' Ph.D. dissertation,
  King's College, Cambridge, UK, May 1989. [Online]. Available:
  \url{http://www.cs.rhul.ac.uk/~chrisw/new_thesis.pdf}
\BIBentrySTDinterwordspacing

\bibitem{Qfit96}
G.~J. Gordon, ``{Stable fitted reinforcement learning},'' in \emph{Advances in
  neural information processing systems (NIPS)}, 1996, p. 1052–1058.

\bibitem{nfq}
M.~Riedmiller, ``{Neural Fitted Q Iteration -- First Experiences with a Data
  Efficient Neural Reinforcement Learning Method},'' in \emph{Proc. ECML},
  2005, pp. 317--328.

\bibitem{ddqn}
H.~v. Hasselt, A.~Guez, and D.~Silver, ``{Deep Reinforcement Learning with
  Double Q-Learning},'' in \emph{Proc.of the Thirtieth AAAI Conference on
  Artificial Intelligence}, 2016, p. 2094–2100.

\bibitem{dueln}
Z.~Wang, T.~Schaul, M.~Hessel, H.~Van~Hasselt, M.~Lanctot, and N.~De~Freitas,
  ``Dueling network architectures for deep reinforcement learning,'' in
  \emph{Proc. of International Conference on Machine Learning}, 2016, p.
  1995–2003.

\bibitem{ac2016}
V.~Mnih, A.~P. Badia, M.~Mirza, A.~Graves, T.~Harley, T.~P. Lillicrap,
  D.~Silver, and K.~Kavukcuoglu, ``Asynchronous methods for deep reinforcement
  learning,'' in \emph{Proc.of the 33rd International Conference on
  International Conference on Machine Learning}, 2016, p. 1928–1937.

\bibitem{ac2018}
A.~Gruslys, W.~Dabney, M.~G. Azar, B.~Piot, M.~Bellemare, and R.~Munos, ``The
  reactor: A fast and sample-efficient actor-critic agent for reinforcement
  learning,'' in \emph{International Conference on Learning Representations
  (ICLR)}, 2018.

\bibitem{ENI:doc005}
``{ETSI -- Experiential Networked Intelligence (ENI): System Architecture},''
  \url{https://www.etsi.org/deliver/etsi_gs/ENI/001_099/005/01.01.01_60/gs_ENI005v010101p.pdf},
  Sep. 2019.

\bibitem{eni:poc}
``{ETSI ENI: Ongoing PoCs},''
  \url{https://eniwiki.etsi.org/index.php?title=Ongoing_PoCs}, Mar. 2021.

\bibitem{Arista}
\BIBentryALTinterwordspacing
``Arista cognitive campus network,'' Jun. 2018. [Online]. Available:
  \url{https://www.arista.com/assets/data/pdf/Whitepapers/Cognitive-Campus-WP.pdf}
\BIBentrySTDinterwordspacing

\bibitem{Ciena}
\BIBentryALTinterwordspacing
``{Ciena Blueplanet: Making Intelligent Automation a Reality with Advanced
  Analytics and Machine Learning},'' May 2019. [Online]. Available:
  \url{https://media.ciena.com/documents/Blue_Planet_Making_Intelligent_Automation_a_Reality_with_Advanced_Analytics_and_Machine_Learning_WP.pdf}
\BIBentrySTDinterwordspacing

\bibitem{Cisco}
\BIBentryALTinterwordspacing
``{Cisco Networks: AI and Machine Learning},'' Mar. 2021. [Online]. Available:
  \url{https://www.cisco.com/c/en/us/solutions/collateral/enterprise-networks/nb-06-ai-nw-analytics-wp-cte-en.html}
\BIBentrySTDinterwordspacing

\bibitem{Ericsson}
\BIBentryALTinterwordspacing
``Ericsson: Artificial intelligence and machine learning in next-generation
  systems,'' Jun. 2018. [Online]. Available:
  \url{https://www.ericsson.com/en/reports-and-papers/white-papers/machine-intelligence}
\BIBentrySTDinterwordspacing

\bibitem{ExtremeNets}
\BIBentryALTinterwordspacing
``{Extreme Networks: Machine Learning and Artificial Intelligence},'' Mar.
  2021. [Online]. Available:
  \url{https://www.extremenetworks.com/solution/machine-learning-and-artificial-intelligence}
\BIBentrySTDinterwordspacing

\bibitem{Juniper}
\BIBentryALTinterwordspacing
``{Juniper Networks: Real AI in Networking},'' Mar. 2021. [Online]. Available:
  \url{https://www.juniper.net/us/en/dm/ai-machine-learning}
\BIBentrySTDinterwordspacing

\bibitem{Vmware}
\BIBentryALTinterwordspacing
``Enabling machine learning as a service {(MLAAS)} with {GPU} acceleration
  using {VM}ware v{R}ealize automation,'' Jul. 2018. [Online]. Available:
  \url{https://www.vmware.com/content/dam/digitalmarketing/vmware/en/pdf/solutions/vmware-mlaas-vrealize-automation-white-paper.pdf}
\BIBentrySTDinterwordspacing

\bibitem{Colab}
``{Google Colaboratory},'' \url{https://colab.research.google.com}, Mar. 2021.

\bibitem{Jupyter}
``{Project Jupyter},'' \url{https://jupyter.org}, Mar. 2021.

\bibitem{keras}
``{Keras: Deep learning for humans},'' \url{https://keras.io/}, Mar. 2021.

\bibitem{pytorch}
``{PyTorch: an open source machine learning framework},''
  \url{https://www.pytorch.org}, Mar. 2021.

\bibitem{scikit}
``{scikit-learn: Machine Learning in Python},''
  \url{https://www.scikit-learn.org}, Mar. 2021.

\bibitem{matlab}
``{Machine Learning with MATLAB},''
  \url{https://in.mathworks.com/solutions/machine-learning.html}, Mar. 2021.

\bibitem{gym}
``{Gym Toolkit},'' \url{https://gym.openai.com/}, Mar. 2021.

\bibitem{stablebl}
``{Stable Baselines},'' \url{https://github.com/hill-a/stable-baselines}, Mar.
  2021.

\bibitem{rlblzoo}
``{RL Baselines Zoo},'' \url{https://github.com/araffin/rl-baselines-zoo}, Mar.
  2021.

\bibitem{tensorflow}
``{TensorFlow: an end-to-end open source platform for machine learning},''
  \url{https://www.tensorflow.org/overview}, Mar. 2021.

\bibitem{noms2020}
D.~{Corcoran}, P.~{Kreuger}, and C.~{Schulte}, ``{Efficient Real-Time Traffic
  Generation for 5G RAN},'' in \emph{Proc. of IEEE/IFIP Network Operations and
  Management Symposium (NOMS)}, 2020, pp. 1--9.

\bibitem{GanGoof2016}
I.~Goodfellow, J.~Pouget-Abadie, M.~Mirza, B.~Xu, D.~Warde-Farley, S.~Ozair,
  A.~Courville, and Y.~Bengio, ``Generative adversarial nets,'' \emph{Advances
  in Neural Information Processing Systems}, vol.~3, pp. 2672--2680, 2014.

\bibitem{GANFlow2019}
M.~Ring, D.~Schlor, D.~Landes, and A.~Hotho, ``Flow-based network traffic
  generation using generative adversarial networks,'' \emph{Elsevier Computers
  \& Security}, vol.~82, pp. 156--172, 2019.

\bibitem{GANShare2020}
Z.~Lin, A.~Jain, C.~Wang, G.~Fanti, and V.~Sekar, ``{Using GANs for Sharing
  Networked Time Series Data: Challenges, Initial Promise, and Open
  Questions},'' in \emph{Proc.of the ACM Internet Measurement Conference},
  2020, p. 464–483.

\bibitem{Ref:NueralPacketClassification}
E.~Liang, H.~Zhu, X.~Jin, and I.~Stoica, ``Neural packet classification,'' in
  \emph{Proc.of ACM SIGCOMM}, 2019, p. 256–269.

\bibitem{Ref:Classbench}
D.~E. Taylor and J.~S. Turner, ``Classbench: A packet classification
  benchmark,'' \emph{IEEE/ACM Transactions on Networking}, vol.~15, no.~3, pp.
  499--511, 2007.

\bibitem{Ref:HiCuts}
P.~Gupta and N.~McKeown, ``Packet classification using hierarchical intelligent
  cuttings,'' in \emph{Proc. of Hot Interconnects VII}, vol.~40, 1999.

\bibitem{Ref:HyperCuts}
S.~Singh, F.~Baboescu, G.~Varghese, and J.~Wang, ``Packet classification using
  multidimensional cutting,'' in \emph{Proc.of ACM SIGCOMM}, 2003, p.
  213–224.

\bibitem{Ref:EffiCuts}
B.~Vamanan, G.~Voskuilen, and T.~N. Vijaykumar, ``Efficuts: Optimizing packet
  classification for memory and throughput,'' in \emph{Proc.of ACM SIGCOMM},
  2010, p. 207–218.

\bibitem{Ref:CutSplit}
W.~Li, X.~Li, H.~Li, and G.~Xie, ``Cutsplit: A decision-tree combining cutting
  and splitting for scalable packet classification,'' in \emph{Proc. of IEEE
  INFOCOM}, 2018, pp. 2645--2653.

\bibitem{STEP2020}
L.~{Yu}, M.~{Li}, W.~{Jin}, Y.~{Guo}, Q.~{Wang}, F.~{Yan}, and P.~{Li}, ``Step:
  A spatio-temporal fine-granular user traffic prediction system for cellular
  networks,'' \emph{IEEE Transactions on Mobile Computing (Early Access)}, Jun.
  2020.

\bibitem{GCNSurvey2019}
S.~Zhang, H.~Tong, J.~Xu, and R.~Maciejewski, ``Graph convolutional networks: a
  comprehensive review,'' \emph{Computational Social Networks}, vol.~6, no.~1,
  p.~11, 2019.

\bibitem{GNNSurvey2021}
Z.~{Wu}, S.~{Pan}, F.~{Chen}, G.~{Long}, C.~{Zhang}, and P.~S. {Yu}, ``{A
  Comprehensive Survey on Graph Neural Networks},'' \emph{IEEE Transactions on
  Neural Networks and Learning Systems}, vol.~32, no.~1, pp. 4--24, 2021.

\bibitem{GRU2014}
J.~Chung, C.~Gulcehre, K.~Cho, and Y.~Bengio, ``Empirical evaluation of gated
  recurrent neural networks on sequence modeling,'' in \emph{Proc. of NIPS
  Workshop on Deep Learning}, Dec. 2014.

\bibitem{satam2020wids}
P.~Satam and S.~Hariri, ``{WIDS: An Anomaly Based Intrusion Detection System
  for Wi-Fi (IEEE 802.11) Protocol},'' \emph{IEEE Transactions on Network and
  Service Management}, vol.~18, no.~1, pp. 1077--1091, Mar. 2021.

\bibitem{NBLLRoute2019}
L.~{Li}, Y.~{Zhang}, W.~{Chen}, S.~K. {Bose}, M.~{Zukerman}, and G.~{Shen},
  ``{Naive Bayes Classifier-Assisted Least Loaded Routing for Circuit-Switched
  Networks},'' \emph{IEEE Access}, vol.~7, pp. 11\,854--11\,867, 2019.

\bibitem{DeepCog}
D.~{Bega}, M.~{Gramaglia}, M.~{Fiore}, A.~{Banchs}, and X.~{Costa-Perez},
  ``{DeepCog: Cognitive Network Management in Sliced 5G Networks with Deep
  Learning},'' in \emph{Proc. of IEEE INFOCOM}, 2019, pp. 280--288.

\bibitem{Adam2014}
D.~P. Kingma and J.~Ba, ``Adam: {A} method for stochastic optimization,'' in
  \emph{Proc. of International Conference on Learning Representations (ICLR)},
  San Diego, CA, USA, May 2015.

\bibitem{DeepCogSC}
D.~{Bega}, M.~{Gramaglia}, M.~{Fiore}, A.~{Banchs}, and X.~{Costa-Perez},
  ``Deepcog source code,'' \url{https://github.com/wnlUc3m/deepcog}, 2019.

\end{thebibliography}

\end{document}